\documentclass[final,3p,times]{elsarticle}

\usepackage{lineno,hyperref}
\usepackage{graphicx}
\usepackage{dcolumn}
\usepackage{bm}
\usepackage{float}
\usepackage{epsfig}
\usepackage{booktabs}
\usepackage{subfigure}
\usepackage{graphics}
\usepackage{amssymb}
\usepackage{amsmath}
\usepackage{array}
\usepackage{color}
\usepackage{booktabs}
\usepackage{multirow}
\usepackage{setspace}
\usepackage{nomencl}
\usepackage [latin1]{inputenc}

\makenomenclature

\def\Dt{\Delta t}

\usepackage{graphicx}

\modulolinenumbers[5]

\journal{Computers and Mathematics with Applications}

\begin{document}

\begin{frontmatter}

\title{A block triple-relaxation-time lattice Boltzmann model for nonlinear anisotropic convection-diffusion equations }

\author[myfirstaddress,mymainaddress]{Yong Zhao}
\author[myfirstaddress,mymainaddress]{Yao Wu}
\author[myfirstaddress,mymainaddress]{Zhenhua Chai}

\author[myfirstaddress,mymainaddress]{Baochang Shi\corref{mycorrespondingauthor}}
\cortext[mycorrespondingauthor]{Corresponding author}
\ead{shibc@hust.edu.cn}

\address[myfirstaddress]{Hubei Key Laboratory of Engineering Modeling and Scientific Computing, Huazhong University of Science and Technology, Wuhan
    430074, China}
\address[mymainaddress]{School of Mathematics and Statistics, Huazhong University of Science and Technology, Wuhan 430074, China}

\begin{abstract}
A block triple-relaxation-time (B-TriRT) lattice Boltzmann model for general nonlinear anisotropic convection-diffusion equations (NACDEs) is proposed, and the Chapman-Enskog analysis shows that the present B-TriRT model can recover the NACDEs correctly. There are some striking features of the present B-TriRT model: firstly, the relaxation matrix of B-TriRT model is  partitioned into three relaxation parameter blocks, rather than a diagonal matrix in general multiple-relaxation-time (MRT) model; secondly, based on the analysis of half-way bounce-back (HBB) scheme for Dirichlet boundary conditions, we obtain an expression to determine the relaxation parameters; thirdly, the anisotropic diffusion tensor can be recovered by the relaxation parameter block that corresponds to the first-order moment of non-equilibrium distribution function. A number of simulations of isotropic and anisotropic convection-diffusion equations are conducted to validate the present B-TriRT model. The results indicate that the present model has a second-order accuracy in space, and  is also more accurate and more stable than some available lattice Boltzmann models.
\end{abstract}


\end{frontmatter}


\section{Introduction}

Lattice Boltzmann method (LBM) has now become a powerful numerical approach for simulating fluid flows and complex physical phenomena~\cite{benzi1992lattice,Chen1998LATTICE,succi2001lattice,Aidun2010Lattice,guo2013lattice,kruger2017lattice}. Unlike conventional computational fluid dynamics (CFD) methods based on the macroscopic continuum equations, LBM is a mesoscopic kinetic-based method, and has some distinct advantages in the treatment of complex boundary and parallel computing scalability. In the past decades, LBM has gained a great success in a variety of fields, including the compressible flows~\cite{alexander1992lattice,Sun1998Lattice,gan2018discrete,saadat2019lattice}, heat and mass transfer in porous media{\color{blue}~\cite{pan2006evaluation,wang2007lattice,chai2016comparative,liu2016multiphase,chai2019lattice}}, blood flows~\cite{fang2002lattice,ouared2005lattice,zhang2007immersed,huang2013non}, thermal flows~{\color{blue}\cite{he1998novel,peng2003simplified,Guo2007Thermal,prasianakis2007lattice,WANG2019,zhao2019comparative}}, multicomponent and multiphase flows~\cite{shan1995multicomponent,He1998Discrete,zheng2006lattice,chen2014critical,zhao2018lattice,chai2018comparative,liang2019axisymmetric}, to name but a few. On the other hand, it also shows great potential in the study of  nonlinear problems, such as  reaction-diffusion equation~\cite{ponce1993lattice,wolf1995lattice,Yu2006A,huber2010lattice}, isotropic convection-diffusion equations (CDEs)~\cite{He2000Lattice,Baochang2009Lattice,Chopard2009The,Chai2013Lattice}, anisotropic convection-diffusion equations~\cite{zhang2002lattice,Ginzburg2005Equilibrium,Ginzburg2005Generic,Ginzburg2007Lattice,Ginzburg2012Truncation,Ginzburg2013Multiple,yoshida2010multiple,huang2014modified,chai2016multiple}, and some high-order partial differential equations~\cite{chai2008unified,lai2009lattice,otomo2018efficient,chai2018lattice}.

Actually, the most widely used model for nonlinear problems is lattice Bhatnagar-Gross-Krook (LBGK) model due to its high computational efficiency, but it is usually unstable for the convection-dominated problems~\cite{Aidun2010Lattice}. To overcome this problem, some improved models have been proposed which can be generally grouped into two major categories: (1) the models through introducing additional parameters; (2) the models through modifying collision operator. Based on the time-splitting scheme of Boltzmann equation, Guo et~al.~\cite{Guo2001A}  proposed a general propagation lattice Boltzmann model (GPLBM) for fluid flows. Subsequently, the model is also extended to solve nonlinear CDEs~\cite{Guo2018General}. Compared to LBGK model, GPLBM can improve the numerical stability by properly adjusting two free parameters to make the Courant-Friedricks-Lewey (CFL) number smaller than 1. However, the convergence will become very slow due to the adoption of a small time-step. Recently, Xiang et al.~\cite{XIANG20122415} introduced a tunable parameter $\beta$ to keep the dimensionless relaxation time $\tau$ away from 0.5 such that the stability of LBGK model can be improved. However, it is not convenient to choose a proper $\beta$, and the improvement of stability is not significant. Different from aforementioned models that introduce some additional parameters, a series of regularized lattice Boltzmann models (RLBMs) for fluid dynamics have also been proposed~\cite{latt2006lattice,zhang2006efficient,montessori2014regularized,mattila2017high}. The main idea of RLBM is to regularize the pre-collision distribution functions so as to achieve better accuracy and stability. Actually, as pointed out by Mattila~\cite{mattila2017high}, the regularization in LBM is a Hermite expansion of non-equilibrium distribution functions, which is used to filter the high-order non-equilibrium moments that make nonhydrodynamic contributions, so that the stability of LBM is improved. Following the similar way, Wang~\cite{Wang2015Regularized} extended the RLBM to solve nonlinear CDEs, and the results show that RLBM is more stable than traditional LBGK models. Afterward, Wang et~al. also applied RLBM to investigate the fluid flows coupled with CDEs, and found that the RLBM can be written as an MRT version in \cite{wang2016regularized}. In another attempt to suppress instabilities, Inamuro~\cite{Inamuro2002A} proposed a lattice kinetic scheme (LKS) for incompressible fluid flows. In the LKS, a gradient term related to the shear rate or temperature gradient is added in the equilibrium distribution function to make the relaxation time to be unity, thus the stability of the LBM can be improved. Up to now, the LKS has been extended to some different fields~\cite{Inamuro2006two,Peng2004thermal,YOSHINO200769,NISHIYAMA2013395}. It should be noted that, the original LKS does not satisfy mass conservation law, and also destroys the localization of collision process since a finite-difference scheme is used to calculate the gradient term. To address these defects, Yang et~al.~\cite{Yang2014Generalized} proposed a generalized modification lattice Boltzmann model (MLBM) for incompressible fluid flows coupled with CDEs. Different from the original LKS, Yang put the gradient term into the source term rather than equilibrium distribution function. Almost at the same time, Wang~\cite{wang_mi_meng_guo_2015} et~al. also proposed a modified version of LKS to guarantee the local mass conservation, and extended it to study non-newtonian fluid flows. Then, Zhao et~al.~\cite{ZHAO20181570} analyzed above two-dimensional modified version of LKS from a mathematical point of view, and found that the modified scheme is essentially a two-relaxation-time (TRT) model. More recently, based on Chai's work \cite{chai2014nonequilibrium}, Wang et al. proposed a modified LKS~\cite{wang2018lattice} for thermal flows. In the work, Wang et al. claimed that ``the convection-diffusion equation (CDE) without the source term can be recovered with a deviation term" in MLBM, and also that ``Although this deviation term can be neglected under some assumptions $\cdots$, it still has an influence on the accuracy of the LB mode". We would like to point out that the these statements of MLBM are misunderstandings. The reason is that when the CDE is coupled with the flow field, the so-called ``deviation term" in MLBM (Ref. \cite{Yang2014Generalized}, Eq. (A14)): $\partial_{t1}\phi\textbf{u}+\lambda\nabla_1\cdot(\phi\textbf{u}\textbf{u})$ can be transformed to $\phi\textbf{a}^{(1)}$ with a $O(Ma^2)$ term being neglected reasonably. On the other hand, the treatment of the unwanted deviation term and the neglected term in Wang's modified LKS are consistent with those in MLBM, and the numerical results presented in Ref.~\cite{wang2018lattice} also confirmed that. In addition, following the previous work \cite{ZHAO20181570}, Wang et al. also performed a matrix analysis to demonstrate that both the RLBM and MLBM are TRT model essentially. After a review on the RLBM and MLBM, we would like to give some remarks: firstly, the RLBM and MLBM share similar features, but they are not included in each other; secondly, the analysis performed by Zhao et~al.~\cite{ZHAO20181570} and Wang et~al.~\cite{wang2018lattice} are limited to two-dimensional case. In this work, we will propose a new model where RLBM and MLBM are its special cases, and extend the analysis in Ref.~\cite{ZHAO20181570,wang2018lattice} to higher-dimensional case. Furthermore, to improve the accuracy of present model, we expand the non-equilibrium distribution function to the second-order moment by Hermite polynomial, and assign a separate relaxation parameters to each Hermite components. We would like to point out the similar idea is also presented in Ref.~\cite{shan2018central}. Through a detailed matrix analysis, we demonstrated that present model can be written as a MRT form, and the relaxation parameter matrix can be partitioned into three relaxation parameter blocks. Finally, based on the analysis of HBB scheme for Dirichlet boundary conditions, we derived an expression to determine the relaxation parameters.

In addition, the  LBGK models are usually limited to  the isotropic CDEs since it is difficult to directly describe the anisotropic diffusion in NACDEs. To overcome the inherent defect in LBGK models, the TRT and MRT models for NACDEs are proposed by Ginzburg  ~\cite{Ginzburg2005Equilibrium,Ginzburg2005Generic,Ginzburg2007Lattice,Ginzburg2012Truncation,Ginzburg2013Multiple}. However, some assumptions made in Ginzburg's models may be not satisfied for some special NACDEs~\cite{Baochang2009Lattice}. Recently, Chai~et~al.\cite{chai2016multiple} developed an MRT lattice Boltzmann model, and the general NACDEs can be correctly recovered. Although the MRT model has the ability to solve the anisotropic problems, it is not convenient to determine the free relaxation parameters, and additionally  multiple-relaxation collision gives rise to more computational cost. Moreover, we also note that the anisotropic diffusion tensor is related to the first-order moment of non-equilibrium distribution function, and thus the corresponding relaxation parameter should be replaced by a matrix to describe the anisotropic diffusion tensor in NACDEs. For this reason, we utilize the relaxation parameter block that corresponds to the first-order moment of non-equilibrium distribution function to recover the anisotropic diffusion tensor.

The paper is organized as follows. In Sect. 2, the B-TriRT model for NACDEs is first presented, then a detailed Chapman-Enskog analysis is conducted. In Sect. 3, we performed a analysis of HBB scheme for Dirichlet boundary conditions to derive the relational expression to determine relaxation parameters. In Sect. 4, several numerical simulations are performed to test the accuracy and stability of present B-TriRT model, and finally some conclusions are summarized in Sect. 5. In addition, the matrix analysis is presented in Appendix.

\section{The block triple-relaxation-time lattice Boltzmann model}
As discussed in the Introduction, the RLBM and MLBM are developed from different points of view, but both of them can be written in a two-relaxation-time version. In this section, we will develop a B-TriRT lattice Boltzmann model, in which the RLBM and MLBM are its special cases. To see the relation between the present B-TriRT model, RLBM and MLBM, we will first present a brief introduction to the RLBM and MLBM, and then develop a B-TriRT lattice Boltzmann model.
\subsection{Regularized lattice Boltzmann model}
In the standard LBM, the equilibrium  distribution function is projected onto a truncated Hilbert subspace $\mathcal{H}_q$ which is spanned by a series of Hermite polynomials, while the non-equilibrium part is not. The regularization procedure work this out by projecting the relevant non-equilibrium moments onto the same subspace $\mathcal{H}_q$ while filtering out the nonhydrodynamic moments ~\cite{mattila2017high}. With the aid of regularization procedure, both the equilibrium and non-equilibrium effects are limited to the subspace so that the stability of model is improved.

The evolution equation of regularized lattice Boltzmann model for CDEs reads ~\cite{Wang2015Regularized}
\begin{equation}\label{eq2}
f_i(\textbf{x}+\textbf{c}_i\Delta t,t+\Delta t) = f_i(\textbf{x},t) -f^{neq}_i(\textbf{x},t)- \left(\frac{1}{\tau}-1\right)\frac{\omega_i \textbf{c}_i\cdot \sum_j \textbf{c}_jf^{neq}_j}{c^2_s}+\Delta t G_i(\textbf{x},t)+\Delta t S_i(\textbf{x},t)+\frac{\Delta t^2}{2}D_iS_i(\textbf{x},t),
\end{equation}
where $f_i(\textbf{x},t),$ and $f_i^{eq}(\textbf{x},t)$ represent the distribution function and  equilibrium distribution function with the discrete velocity $\textbf{c}_i$ at time $t$ in location $\textbf{x}$ respectively, $f_i^{neq}=f_i-f_i^{eq}$ is the non-equilibrium distribution function, $\tau$ is the dimensionless relaxation time, $\omega_i$ are the weight coefficients, $c_s$ is the lattice sound velocity, $G_i(\textbf{x},\phi,t)$ is the correction term to eliminate error caused by the convection term, $S_i(\textbf{x},\phi,t)$ is {\color{blue}the distribution function of source term}, $D_i=\partial_t + \textbf{c}_i\cdot\nabla$ is {\color{blue}the gradient operator containing time and space derivatives}.

\subsection{Modified lattice Boltzmann model}
The MLBM originates from LKS which is first proposed by Inamuro~\cite{Inamuro2002A}. The main idea of  LKS or MLBM is to introduce a gradient term  in the  equilibrium distribution function or source term to keep the relaxation time in a proper range. In this work, we mainly focus on the MLBM, and its  evolution function for CDEs can be  written as
\begin{equation}
f_i(\textbf{x}+\textbf{c}_i\Delta t,t+\Delta t)=f_i-\frac{1}{\tau}f^{neq}_i(\textbf{x},t)-\left(\frac{1}{\tau_Z}-\frac{1}{\tau}\right)\frac{\omega_i \textbf{c}_i\cdot \sum_j \textbf{c}_jf^{neq}_j}{c^2_s}+\Delta t G_i+\Delta t S_i+\frac{\Delta t^2}{2}D_iS_i,
\label{eq3}
\end{equation}
where $\tau_Z=\tau-Z$, $Z$ is a tunable parameter.

\subsection{The block triple-relaxation-time lattice Boltzmann model}

Based on Eqs.~(\ref{eq2}) and (\ref{eq3}), one can easily find that both RLBM and MLBM are specific form of the following unified evolution equation,

\begin{equation}\label{eq4}
f_i(\textbf{x}+\textbf{c}_i\Delta t,t+\Delta t)=f_i-k_0f^{neq}_i(\textbf{x},t)-(k_1-k_0)\frac{\omega_i \textbf{c}_i\cdot \textbf{M}_1^{neq}(\textbf{x},t)}{c^2_s}+\Delta t G_i+\Delta t S_i+\frac{\Delta t^2}{2}D_iS_i,
\end{equation}
where $\textbf{M}_1^{neq}(\textbf{x},t)=\sum_j \textbf{c}_jf^{neq}_j$ is the first-order moment of non-equilibrium distribution function $f^{neq}_j$, where $k_0, k_1 \in (0,2)$ are dimensionless relaxation parameters.

In addition, in order to improve accuracy of the LB model Eq.~(\ref{eq4}), a natural idea is to expand the non-equilibrium distribution function to higher-order moments by Hermite polynomial. However, for the third-order and higher order moments, a multi-velocity lattice model is required. Hence, we only retain the second-order moment of $f_i^{neq}$. Then, the evolution equation of present lattice Boltzmann model reads,
\begin{equation}
\begin{aligned}\label{eq5}
f_i(\textbf{x}+\textbf{c}_i\Delta t,t+\Delta t)=& f_i(\textbf{x},t)-k_0f_i^{neq}(\textbf{x},t)-(k_1-k_0)\frac{\omega_i\textbf{c}_i\cdot \textbf{M}_1^{neq}(\textbf{x},t)}{c_s^2}-(k_2-k_0)\frac{\omega_i(\textbf{c}_i\textbf{c}_i-c_s^2\textbf{I}):\textbf{M}_2^{neq}(\textbf{x},t)}{2c_s^4}\\
&+\Delta tG_i(\textbf{x},t)+\Delta t S_i(\textbf{x},t)+\frac{\Delta t^2}{2}\bar{D}_i S_i(\textbf{x},t),
\end{aligned}
\end{equation}
where $\textbf{M}_2^{neq}(\textbf{x},t)=\sum_j \textbf{c}_j \textbf{c}_j f^{neq}_j$ is the second-order moment of non-equilibrium distribution function $f^{neq}_j$, and $\bar{D}_i=\partial_t + \gamma\textbf{c}_i\cdot \nabla$ with $\gamma$ being a tunable parameter to be determined in the following part. Furthermore, to solve some more complicated problems, (e.g., the anisotropic problems), we set the relaxation parameters $k_1$ and $k_2$ as  matrices and rewrite Eq.~(\ref{eq5}) as
\begin{equation}
\begin{aligned}\label{eq6}
f_i(\textbf{x}+\textbf{c}_i\Delta t,t+\Delta t)=& f_i(\textbf{x},t)-k_0f_i^{neq}(\textbf{x},t)-\frac{\omega_i\textbf{c}_i\cdot (\textbf{K}_1-k_0\textbf{I}) \textbf{M}_1^{neq}(\textbf{x},t)}{c_s^2}-\frac{\omega_i(\textbf{c}_i\textbf{c}_i-c_s^2\textbf{I}):(\textbf{K}_2-k_0\hat{\textbf{I}})\circ\textbf{M}_2^{neq}(\textbf{x},t)}{2c_s^4}\\
&+\Delta tG_i(\textbf{x},t)+\Delta t S_i(\textbf{x},t)+\frac{\Delta t^2}{2}\bar{D}_i S_i(\textbf{x},t),
\end{aligned}
\end{equation}
where the $\textbf{K}_1$ and $\textbf{K}_2$ are $d \times d$ invertible and positive matrices {\color{blue} respectively,}  with $d$ being the spatial dimension, $\textbf{I}$ is an indentity matrix, $\hat{\textbf{I}}$ is a matrix with $\hat{\textbf{I}}_{ij}=1,~(\forall~i,~j)$, $\circ$ represents the Hadamard product.

In the following, we will list some remarks on the present model:\\
$Remark ~\uppercase\expandafter{\romannumeral1}:$ Based on a careful matrix analysis in Appendix, we can show that the evolution equation of present B-TriRT model can be written as an MRT form,
\begin{equation}\label{eqq}
  f_i(\textbf{x}+\textbf{c}_i\Delta t,t+\Delta t)=f_i(\textbf{x},t)-(\textbf{T}^{-1}\textbf{S}_f\textbf{T})_{ij}f_j^{neq}+\Dt G_i+\Dt S_i + \frac{\Dt^2}{2}\bar{D}_iS_i,
\end{equation}
where $\textbf{T}$ is the transformation matrix, and $\textbf{S}_f$ is the relaxation matrix, both of them are given in Appendix. In addition, we also find that the relaxation matrix $\textbf{S}_f$ could be partitioned into several relaxation parameter blocks as,
\begin{equation}\label{eqqq}
 \textbf{S}_f=\mathrm{diag}(\textbf{S}_0, \textbf{S}_1, \textbf{S}_2, \cdots, \textbf{S}_m),~~~~ m<q,
\end{equation}
where $m$ represents the order of moment, $q$ represents the number of discrete velocities. Based on the concept of block, the general multiple-relaxation-time could be extended to a block multiple-relaxation-time version. Actually, as mentioned above, for the third-order and higher-order moments, a multi-velocity model is required. For this reason, we only partition the relaxation matrix into three relaxation parameter blocks as: $\textbf{S}_0,~\textbf{S}_1$ and $\textbf{S}_2$, the model is also named B-TriRT model, where $\textbf{S}_1=\textbf{K}_1$ that corresponds to the first-order moment of non-equilibrium distribution function, $\textbf{S}_2$ is related to $\textbf{K}_2$ that corresponds to second-order moment of non-equilibrium distribution function, $\textbf{S}_0=k_0\textbf{I}$ that corresponds to the remaining moments of non-equilibrium distribution function. In addition, the present B-TriRT model can also be reduced to TRT model in Ref.~\cite{ginzburg2008two,ginzburg2008study} by setting $\textbf{K}_2=k_0\hat{\textbf{I}}$ for the $DdQq, q=2d ~or ~2d+1$ lattice model. However, for general $DdQq,~q\neq 2d ~or ~2d+1$ lattice model, the TRT model cannot be categorized into the present B-TriRT model.

$Remark ~\uppercase\expandafter{\romannumeral2}:$ The RLBM and MLBM are special cases of the present B-TriRT model. By adopting the relaxation parameters $\textbf{K}_1=k_1\textbf{I},~\textbf{K}_2=k_2\hat{\textbf{I}}$, and choosing appropriate parameters $k_1,~k_2$, we can obtain the LBGK model, RLBM and MLBM from present model, as seen from Tab.~\ref{tab1}.

 \begin{table}[ht]
 \centering
 \caption{Parameter values corresponding to different models}\label{tab1}
\begin{tabular}{c|c|c|c}
  \hline\hline
  Model & $k_0$ & $k_1$ & $k_2$ \\
   \hline
  LBGK & $1/\tau$ & $1/\tau $& $1/\tau $\\
  MLBM & $1/\tau$& $1/\tau_Z$ & $1/\tau$ \\
  RLBM & 1 & $1/\tau$ & 1 \\
  \hline \hline
\end{tabular}
\end{table}

$Remark ~\uppercase\expandafter{\romannumeral3}:$ As pointed out by Chai~et~al.~\cite{chai2016multiple}, there are two special schemes to treat the source term derivative $\bar{D}_i S_i(\textbf{x},t)$ in evolution equation based on the choice of the parameter $\gamma$.

$\mathbf{Schemes~1}  ~(\gamma=1):$ In this scheme, both the time and space derivatives are contained in the evolution equation. However, when we use the finite-difference scheme to calculate the space derivative, the collision process cannot be conducted locally. Although, we can adopt some spacial technologies as mentioned in \cite{chai2016multiple} to maintain the locality of computation, it brings more matrix multiplication which will reduce the computational efficiency.

$\mathbf{Schemes~2}  ~(\gamma=0):$ In the second scheme, the derivative term $\bar{D}_i S_i(\textbf{x},t)$ would reduce to  $\partial_t S_i(\textbf{x},t)$, and an explicit finite-difference scheme (i.e., $\partial_t S_i(\textbf{x},t)=[S_i(\textbf{x},t)-S_i(\textbf{x},t-\delta t)]/\delta t$) is applied to compute the time derivative. In the present model, we adopt the second scheme ($\gamma=0$) to compute the source term derivative $\bar{D}_i S_i(\textbf{x},t)$ for simplicity, and it can also be implemented locally.

$Remark ~\uppercase\expandafter{\romannumeral4}:$ We would like to point out that the present B-TriRT model  can be adopted to solve both NACDEs and Navier-Stokes equations (NSEs). For NACDEs, the relaxation parameter $\textbf{K}_1$ is related to the diffusion term, while for NSEs, $\textbf{K}_1$ can be set arbitrary since the first-order moment of non-equilibrium distribution function is equal to zero based on momentum conservation law, and the relaxation parameter $\textbf{K}_2$ is related to the viscous term. However, in this work, we mainly focus on the NACDEs,

\begin{equation}\label{eq1}
   \partial_t \phi+\nabla \cdot\textbf{B}(\textbf{x},\phi,t)=\nabla\cdot[\textbf{A}(\textbf{x},\phi,t)\nabla\cdot \textbf{D}(\textbf{x},\phi,t)]+ S(\textbf{x},\phi,t),
\end{equation}
 where $\phi$ is the scalar variable and is a function of space \textbf{x} and time $t$, $\nabla$ is the gradient operator, $\textbf{B}(\textbf{x},\phi,t), \textbf{D}(\textbf{x},\phi,t) $ and $\textbf{S}(\textbf{x},\phi,t)$ are the known convection term, diffusion term and source term. $\textbf{A}(\textbf{x},t)$ is the diffusion tensor, it can also be the function of scalar variable $\phi$, space \textbf{x} and time $t$.

\subsection{The B-TriRT model for NACDEs}
In the B-TriRT model for NACDEs, the equilibrium distribution function $f_i^{eq}(\textbf{x},t)$ is given as
\begin{equation}
f_i^{eq}(\textbf{x},t)=\omega_i\left[\phi + \frac{\textbf{c}_i\cdot \textbf{B}(\textbf{x},\phi,t)}{c_s^2}+\frac{(\textbf{D}(\textbf{x},\phi,t)-\phi \textbf{I}):(\textbf{c}_i\textbf{c}_i-c_s^2\textbf{I})}{2c_s^2} \right],
\label{eq7}
\end{equation}
and the {\color{blue} distribution function of  source term reads}
\begin{equation}
S_i^{eq}(\textbf{x},t)=\omega_i\textbf{S}(\textbf{x},t).
\label{eq8}
\end{equation}
Based on Eqs.~(\ref{eq7},~\ref{eq8}), one can obtain the following conditions,

\begin{equation}
\sum_i f_i=\sum_i f_i^{eq}=\phi,~ \sum_i \textbf{c}_if_i^{eq}=\textbf{B}(\textbf{x},\phi,t),~ \sum_i \textbf{c}_i\textbf{c}_if_i^{eq}=c_s^2\textbf{D}(\textbf{x},\phi,t),
\label{eq9}
\end{equation}

\begin{equation}
\sum_i S_i=S(\textbf{x},\phi,t),~ \sum_i \textbf{c}_iS_i=0.
\label{eq10}
\end{equation}
The correction term $G_i(\textbf{x},\phi,t)$ in Eq.(~\ref{eq6}) is used to eliminate the additional term caused by convection term (i.e.,$\nabla \cdot \partial_t \textbf{B}(\textbf{x},\phi,t),$ see the details in the Chapman-Enskog analysis), and  is defined as
\begin{equation}
 G_i(\textbf{x},\phi,t)=\frac{\omega_i\textbf{c}_i\cdot\left(\textbf{I}-\frac{\textbf{K}_1}{2}\right)\partial_t\textbf{B}(\textbf{x},\phi,t)}{c^2_s}.
\label{eq11}
\end{equation}

In the following part, we will perform a detailed Chapman-Enskog analysis to recover the macroscopic equation~(\ref{eq1}) from evolution equation~(\ref{eq6}). Firstly, we expand the distribution function, the time and space derivatives as

\begin{equation}
\begin{aligned}
&f_i=f_i^{(0)}+\varepsilon f_i^{(1)}+ \varepsilon^2 f_i^{(2)}+\cdots,\\
&\textbf{M}_{r}^{neq}=\textbf{M}_{r}^{neq(0)}+\varepsilon \textbf{M}_{r}^{neq(1)}+\varepsilon^2 \textbf{M}_{r}^{neq(2)}, ~~~~r=1,2,\\
&\textbf{M}_1^{neq(0)}=\sum_i \textbf{c}_i(f_i^{(0)}-f_i^{eq}),\\
&\textbf{M}_2^{neq(0)}=\sum_i \textbf{c}_i\textbf{c}_i(f_i^{(0)}-f_i^{eq}),\\
&\textbf{M}_1^{neq(l)}=\sum_i \textbf{c}_if_i^{(l)},~~~~l=1,2\\
&\textbf{M}_2^{neq(l)}=\sum_i \textbf{c}_i\textbf{c}_if_i^{(l)},~~~~l=1,2\\
&S_i=\varepsilon S_i^{(1)},\\
&G_i=\varepsilon G_i^{(1)}+ \varepsilon^2 G_i^{(2)},\\
&\partial_t=\varepsilon \partial_{t1}+\varepsilon^2 \partial_{t2},\\
&\nabla=\varepsilon \nabla_1.
\end{aligned}
\label{eq12}
\end{equation}

Applying the Taylor series expansion to the evolution equation Eq.~(\ref{eq6}) at time $t$ and position $\textbf{x}$, we have
\begin{equation}
\begin{aligned}
\Delta t D_i f_i+\frac{\Delta^2}{2}D^2_i f_i + \cdots =& -k_0 f_i^{neq}-\frac{\omega_i\textbf{c}_i\cdot (\textbf{K}_1-k_0\textbf{I})\textbf{M}_1^{neq}(\textbf{x},t)}{c_s^2}-\frac{\omega_i(\textbf{c}_i\textbf{c}_i-c_s^2I):(\textbf{K}_2-k_0\hat{\textbf{I}})\circ\textbf{M}_2^{neq}(\textbf{x},t)}{2c_s^4}\\
&+\Delta tG_i(\textbf{x},t)+\Delta t S_i(\textbf{x},t)+\frac{\Delta t^2}{2}\bar{D}_i S_i(\textbf{x},t),
\end{aligned}
\label{eq13}
\end{equation}
Substituting Eq.~(\ref{eq12}) into Eq.~(\ref{eq13}), one can obtain:

\begin{equation}
\begin{aligned}
\Delta t&(\varepsilon D_{1i}+\varepsilon^2\partial_{t2})(f_i^{(0)}+\varepsilon f_i^{(1)} + \varepsilon^2 f_i^{(2)})+ \frac{\Delta t^2}{2}(\varepsilon D_{1i}+\varepsilon^2\partial_{t2})^2(f_i^{(0)}+\varepsilon f_i^{(1)}+ \varepsilon^2 f_i^{(2)})\\
=&-k_0(f_i^{(0)}+\varepsilon f_i^{(1)} + \varepsilon^2 f_i^{(2)}-f_i^{eq})- \frac{\omega_i\textbf{c}_i\cdot (\textbf{K}_1-k_0\textbf{I})(\textbf{M}_{1}^{neq(0)}+\varepsilon \textbf{M}_{1}^{neq(1)}+\varepsilon^2 \textbf{M}_{1}^{neq(2)})}{c_s^2}\\
&-\frac{\omega_i(\textbf{c}_i\textbf{c}_i-c_s^2I): (\textbf{K}_2-k_0\hat{\textbf{I}})\circ(\textbf{M}_{2}^{neq(0)}+\varepsilon \textbf{M}_{2}^{neq(1)}+\varepsilon^2 \textbf{M}_{2}^{neq(2)})}{2c_s^4} +\Delta t(\varepsilon G_i^{(1)}+ \varepsilon^2 G_i^{(2)})+ \Delta t \varepsilon S_i^{(1)}+\frac{\Delta t^2}{2}(\varepsilon^2\bar{D}_{1i}+\varepsilon^3\partial_{t2})S_i^{(1)}.
\end{aligned}
\label{eq14}
\end{equation}
where $D_{1i}=\partial _{t1}+ \textbf{c}_i \cdot \nabla_1, \bar{D}_{1i}=\partial _{t1}+ \gamma\textbf{c}_i \cdot \nabla_1$.

Based on Eq. (\ref{eq14}), one can derive the following equations at different orders of $\varepsilon$:

\begin{eqnarray}
&O(\varepsilon^0): \quad k_0(f_i^{(0)}-f_i^{eq})+\frac{\omega_i\textbf{c}_i\cdot (\textbf{K}_1-k_0\textbf{I})\textbf{M}_1^{neq(0)}}{c_s^2}-\frac{\omega_i(\textbf{c}_i\textbf{c}_i-c_s^2I):(\textbf{K}_2-k_0\hat{\textbf{I}})\circ\textbf{M}_2^{neq(0)}}{2c_s^4}=0, \label{eq15}\\
&O(\varepsilon^1): \quad \Delta t D_{1i}f_i^{(0)}=-k_0f_i^{(1)}-\frac{\omega_i\textbf{c}_i\cdot (\textbf{K}_1-k_0\textbf{I})\textbf{M}_1^{neq(1)}}{c_s^2}-\frac{\omega_i(\textbf{c}_i\textbf{c}_i-c_s^2I):(\textbf{K}_2-k_0\hat{\textbf{I}})\circ\textbf{M}_2^{neq(1)}}{2c_s^4}+\Delta tS_i^{(1)}+\Delta tG_i^{(1)},\label{eq16}\\
&O(\varepsilon^2):  \Delta t D_{1i}f_i^{(1)}+ \Delta t \partial_{t2}f_i^{(0)}+\frac{\Delta t^2}{2}D_{1i}^2 f_i^{(0)}=-k_0f_i^{(2)}-\frac{\omega_i\textbf{c}_i\cdot (\textbf{K}_1-k_0\textbf{I})\textbf{M}_1^{neq(2)}}{c_s^2}-\frac{\omega_i(\textbf{c}_i\textbf{c}_i-c_s^2I):(\textbf{K}_2-k_0\hat{\textbf{I}})\circ\textbf{M}_2^{neq(2)}}{2c_s^4}+\Delta t G_i^{(2)}+ \frac{\Delta t^2}{2}\bar{D}_{1i}S_i^{(1)}. \label{eq17}
\end{eqnarray}

Multiplying Eq.~(\ref{eq15}) by $\textbf{c}_i$ and $\textbf{c}_i\textbf{c}_i$, then summing them over $i$, we get
\begin{equation}\label{eq18}
\textbf{K}_1\textbf{M}_1^{neq(0)}=0,~~~~\textbf{K}_2\circ\textbf{M}_2^{neq(0)} =0,
\end{equation}
based on the fact {\color{blue} that $\textbf{K}_1$ is invertible  and $\textbf{K}_2$  consists of non-zero elements, }one can obtain the following conclusion:
\begin{equation}\label{eq19}
 \textbf{M}_1^{neq(0)} =0,~~~~\textbf{M}_2^{neq(0)}=0.
\end{equation}

Substituting Eq.~(\ref{eq19}) into Eq.~(\ref{eq15}) yields

\begin{equation}\label{eq20}
  k_0(f_i^{(0)}-f_i^{eq})=0,
\end{equation}
which also leads to
\begin{equation}\label{eq21}
 f_i^{(0)}=f_i^{eq}.
\end{equation}

Then from Eqs. (\ref{eq9}), (\ref{eq12}) and (\ref{eq21}), one can easily obtain
\begin{equation}\label{eq22}
 \sum_i f_i^{(n)}=0, n\geq1.
\end{equation}
With the help of Eq.~(\ref{eq16}), we can rewrite the Eq.~(\ref{eq17}) as
\begin{equation}\label{eq23}
\begin{aligned}
 &\Delta t D_{1i}f_i^{(1)}+\Delta t \partial_{t2}f_i^{(0)}+\frac{\Delta t}{2}D_{1i}\left(-k_0f_i^{(1)}-\frac{\omega_i\textbf{c}_i\cdot (\textbf{K}_1-k_0\textbf{I})\textbf{M}_1^{neq(1)}}{c_s^2}-\frac{\omega_i(\textbf{c}_i\textbf{c}_i-c_s^2I):(\textbf{K}_2-k_0\hat{\textbf{I}})\circ\textbf{M}_2^{neq(1)}}{2c_s^4}+\Delta t G_i^{(1)}+ \Delta t S_i^{(1)}\right)\\
 &=-k_0f_i^{(2))}-\frac{\omega_i\textbf{c}_i\cdot (\textbf{K}_1-k_0\textbf{I})\textbf{M}_1^{neq(2)}}{c_s^2}-\frac{\omega_i(\textbf{c}_i\textbf{c}_i-c_s^2\textbf{I}):(\textbf{K}_2-k_0\hat{\textbf{I}})\circ\textbf{M}_2^{neq(2)}}{2c_s^4}+\Delta t G_i^{(2)}+ \frac{\Delta t^2}{2}\bar{D}_{1i}S_i^{(1)}.
 \end{aligned}
\end{equation}
Rearrange the Eq.(\ref{eq23}), we have
\begin{equation}\label{eq24}
\begin{aligned}
 &\Delta t D_{1i}\left(1-\frac{k_0}{2}\right)f_i^{(1)}+\Delta t \partial_{t2}f_i^{(0)}-\frac{\Delta t}{2}D_{1i}\left(\frac{\omega_i\textbf{c}_i\cdot (\textbf{K}_1-k_0\textbf{I})\textbf{M}_1^{neq(1)}}{c_s^2}+\frac{\omega_i(\textbf{c}_i\textbf{c}_i-c_s^2I):(\textbf{K}_2-k_0\hat{\textbf{I}})\circ\textbf{M}_2^{neq(1)}}{2c_s^4}-\Delta tG_i^{(1)}\right)\\
 &=-k_0f_i^{(2)}-\frac{\omega_i\textbf{c}_i\cdot (\textbf{K}_1-k_0\textbf{I})\textbf{M}_1^{neq(2)}}{c_s^2}-\frac{\omega_i(\textbf{c}_i\textbf{c}_i-c_s^2I):(\textbf{K}_2-k_0\hat{\textbf{I}})\circ\textbf{M}_2^{neq(2)}}{2c_s^4}+\Delta t G_i^{(2)}+\frac{\Delta t^2}{2}(\bar{D}_{1i}-D_{1i})S_i^{(1)}.
 \end{aligned}
\end{equation}

Summing Eqs. (\ref{eq16}) and (\ref{eq24}) over $i$, and utilizing Eqs. (\ref{eq7}), (\ref{eq8}), (\ref{eq9}), (\ref{eq22}), one can derive the following equations,

\begin{eqnarray}
&\partial_{t1}\phi+\nabla_1\cdot\textbf{B}(\textbf{x},\phi,t)=S^{(1)},\\ \label{eq25}
&\partial_{t2}\phi+\nabla_1\cdot\left(\textbf{I}-\frac{\textbf{K}_1}{2}\right)\sum_i\textbf{c}_if_i^{(1)}+\frac{\Delta t}{2}\nabla\cdot\left(\textbf{I}-\frac{\textbf{K}_1}{2}\right)\partial_{t1}\textbf{B}(\textbf{x},\phi,t)=0. \label{eq26}
\end{eqnarray}

Multiplying Eq.(\ref{eq16}) by $\textbf{c}_i$ and summing it over $i$, it is easy to get

\begin{equation}\label{eq27}
\begin{aligned}
 \textbf{K}_1\sum_i\textbf{c}_if_i^{(1)}&=-\Delta t\left(\partial_{t1}\sum_i\textbf{c}_if_i^{(0)}+\nabla_1\cdot\sum_i\textbf{c}_i\textbf{c}_if_i^{(0)} \right)+\Delta t \sum_i\textbf{c}_iG_i^{(1)} \\
 &=-\Delta t\left[\partial_{t1}\textbf{B}(\textbf{x},\phi,t)+\nabla_1\cdot c_s^2\textbf{D}(\textbf{x},\phi,t) \right]+\Delta t\left(\textbf{I}-\frac{\textbf{K}_1}{2}\right)\partial_{t1}\textbf{B}(\textbf{x},\phi,t)\\
 &=-\Delta t\nabla_1\cdot c_s^2\textbf{D}(\textbf{x},\phi,t)- \Delta t\frac{ \textbf{K}_1}{2}\partial_{t1}\textbf{B}(\textbf{x},\phi,t).
\end{aligned}
\end{equation}
Multiplying Eq.(\ref{eq27}) by $\textbf{K}_1^{-1}$, and substituting the result into Eq. (\ref{eq26}) yields
\begin{equation}\label{eq28}
  \partial_{t2}\phi=\nabla_1\cdot\left[c_s^2\left(\textbf{K}_1^{-1}-\frac{1}{2}\textbf{I}\right)\Delta t\nabla_1\cdot\textbf{D}(\textbf{x},\phi,t)\right].
\end{equation}
Applying Eq. (\ref{eq25})$\times \varepsilon +$ Eq. (\ref{eq28})$\times \varepsilon^2$, we can recover the NACDE
\begin{equation}\label{eq29}
 \partial_t \phi+\nabla \cdot\textbf{B}(\textbf{x},\phi,t)=\nabla\cdot[\textbf{A}(\textbf{x},\phi,t)\nabla\cdot \textbf{D}(\textbf{x},\phi,t)]+S(\textbf{x},\phi,t),
\end{equation}
with
\begin{equation}\label{eq30}
 \textbf{A}(\textbf{x},\phi,t)=c_s^2\left(\textbf{K}_1^{-1}-\frac{1}{2}\textbf{I}\right)\Delta t.
\end{equation}

\section{Analysis of HBB scheme for Dirichlet boundary conditions}
In this section, we will conduct a analysis of HBB scheme for Dirichlet boundary conditions to derive a relation between $\textbf{K}_1$ and $\textbf{K}_2$.  Here we take a unidirectional steady diffusion problem as an example, which can be described by~\cite{Cui2016Discrete}
\begin{eqnarray}\label{eq31}
 \partial_t \phi + \nabla \cdot (\phi \textbf{u}) = \nabla \cdot (\alpha\nabla \phi)+S, \\
  \phi(t,x,0)=\phi_0,~~ \phi(t,x,L)=\phi_L,\label{eq32}
\end{eqnarray}
where $\textbf{u}=(u_x,0)^T$ with $u_x$ being a constant, $\alpha$ is a constant diffusion coefficient, $S=2\alpha\Delta \phi/L^2$ is the source term with $\Delta \phi=\phi_L-\phi_0$. One can easily derive its analytical solution,
\begin{equation}\label{eq36}
\phi(x,y)=\phi_0 + \Delta \phi y(2-y).
\end{equation}

To obtain the numerical solution of this problem, we first give the equivalent difference equation of present model. Since this example is an unidirectional steady diffusion problem, and  the source term $S$ is a constant, we can adopt the following linear equilibrium distribution function,
\begin{equation}\label{eq37}
f_i^{eq}(\textbf{x},t)=\omega_i\phi\left(1+\frac{\textbf{c}_i\cdot\textbf{u}}{c_s^2}\right),
\end{equation}
and  the evolution equation is composed of the collision and propagation steps,
\begin{equation}\label{eq38}
{\color{blue}f_i^+(\textbf{x},t)=f_i(\textbf{x},t)-(\textbf{T}^{-1}\textbf{S}_f\textbf{T})_{ij}f_j^{neq}+\Dt \omega_i S,} ~~~~~~~~  f_i(\textbf{x}+\textbf{c}_i\Delta t,t+\Delta t)=f_i^+(\textbf{x},t),
\end{equation}
where the superscript ``+" means post-collision. Here we employ the popular D2Q9 lattice model for the problem and the weight coefficients are given as $\omega_0=4/9, \omega_{1,2,3,4}=1/9, \omega_{5,6,7,8}=1/36$. For this lattice model, we can obtain the following equations from Eqs.~(\ref{eq37})-(\ref{eq38}),

\begin{equation}\label{eq39}
\begin{aligned}
 f^n_0+f^n_1+f^n_3=&f^n_0+f^n_1+f^n_3-k_0\left[f^n_0+f^n_1+f^n_3-(f^{n,eq}_0+f^{n,eq}_1+f^{n,eq}_3)\right]\\
 &-(k_0-k_2)\left[f^n_2+f^n_4+f^n_5+f^n_6+f^n_7+f^n_8-(f^{n,eq}_2+f^{n,eq}_4+f^{n,eq}_5+f^{n,eq}_6+f^{n,eq}_7+f^{n,eq}_8)\right]+\frac{2}{3}\Delta t S,
 \end{aligned}
\end{equation}
\begin{flushleft}
  \begin{equation}\label{eq40}
\begin{aligned}
 f^{n+1}_2+f^{n+1}_5+f^{n+1}_6=&f^n_2+f^n_5+f^n_6-\left(\frac{k_1}{2}+\frac{k_2}{2}\right)\left[f^n_2+f^n_5+f^n_6-(f^{n,eq}_2+f^{n,eq}_5+f^{n,eq}_6)\right]\\
 &-\left(\frac{k_2}{2}-\frac{k_1}{2}\right)\left[f^n_4+f^n_7+f^n_8-(f^{n,eq}_4+f^{n,eq}_7+f^{n,eq}_8)\right]+\frac{1}{3}\Delta t S,
 \end{aligned}
\end{equation}
\end{flushleft}
\begin{equation}\label{eq41}
\begin{aligned}
 f^{n-1}_4+f^{n-1}_7+f^{n-1}_8=&f^n_4+f^n_7+f^n_8-(\frac{k_2}{2}-\frac{k_1}{2})\left[f^n_2+f^n_5+f^n_6-(f^{n,eq}_2+f^{n,eq}_5+f^{n,eq}_6)\right]\\
 &-\left(\frac{k_1}{2}+\frac{k_2}{2}\right)\left[f^n_4+f^n_7+f^n_8-(f^{n,eq}_4+f^{n,eq}_7+f^{n,eq}_8)\right]+\frac{1}{3}\Delta t S,
 \end{aligned}
\end{equation}
where $f^n_i$ is the distribution function at the layer $n$ in the $y$ direction.

Based on the Eqs.~(\ref{eq9}) and (\ref{eq37}) , we can rewrite the Eq.~(\ref{eq39}) as
\begin{equation}\label{eq42}
f^n_0+f^n_1+f^n_3=\frac{2}{3}\phi_n+\frac{2}{3k_2}\Delta tS,
\end{equation}
then according to Eq.~(\ref{eq10}), we also have
\begin{equation}\label{eq43}
(f^n_2+f^n_5+f^n_6)+(f^n_4+f^n_7+f^n_8)=\frac{1}{3}\phi_n-\frac{2}{3k_2}\Delta t S.
\end{equation}

Substituting Eq. (\ref{eq43}) into Eq.~(\ref{eq40}) and Eq.~(\ref{eq41}) yields
\begin{equation}\label{eq44}
 f^{n+1}_2+f^{n+1}_5+f^{n+1}_6=\frac{2-k_1}{6}\phi_n+\frac{k_1}{2c}\phi_nu_{y,n}+(k_1-1)(f^n_4+f^n_7+f^n_8)+\frac{1}{6}\Delta tS-\left(1-\frac{k_1}{2}-\frac{k_2}{2}\right)\frac{2}{3k_2}\Delta t S,
\end{equation}
\begin{equation}\label{eq45}
 f^{n-1}_4+f^{n-1}_7+f^{n-1}_8=\frac{2-k_1}{6}\phi_n+\frac{k_1}{2c}\phi_nu_{y,n}+(k_1-1)(f^n_2+f^n_5+f^n_6)+\frac{1}{6}\Delta tS-\left(1-\frac{k_1}{2}-\frac{k_2}{2}\right)\frac{2}{3k_2}\Delta tS.
\end{equation}

Based on Eqs.~(\ref{eq44}) and (\ref{eq45}), we can obtain the following equations,
\begin{equation}\label{eq46}
 f^{n}_2+f^{n}_5+f^{n}_6=\frac{2-k_1}{6}\phi_{n-1}+(k_1-1)(f^{n-1}_4+f^{n-1}_7+f^{n-1}_8)+\frac{k_1}{2c}\phi_{n-1}u_{y,n-1}+\frac{1}{6}\Delta tS-\left(1-\frac{k_1}{2}-\frac{k_2}{2}\right)\frac{2}{3k_2}\Delta tS
\end{equation}

\begin{equation}\label{eq47}
 f^{n}_4+f^{n}_7+f^{n}_8=\frac{2-k_1}{6}\phi_{n+1}+(k_1-1)(f^{n+1}_2+f^{n+1}_5+f^{n+1}_6)+\frac{k_1}{2c}\phi_{n+1}u_{y,n+1}+\frac{1}{6}\Delta tS-\left(1-\frac{k_1}{2}-\frac{k_2}{2}\right)\frac{2}{3k_2}\Delta tS.
\end{equation}

A summation of Eq.~(\ref{eq46}) and Eq.~(\ref{eq47}), and  with aid of Eqs.~(\ref{eq44}) and (\ref{eq45}), we can derive the equivalent difference equation of present model as
\begin{equation}\label{eq48}
\frac{\phi_{n+1}u_{y,n+1}-\phi_{n-1}u_{y,n-1}}{2\Delta x}=\alpha\frac{\phi_{n-1}-2\phi_{n}+\phi_{n+1}}{\Delta x^2}+S.
\end{equation}
where the diffusion coefficient $\alpha$ is given by Eq.~(\ref{eq30}) with $c_s^2=c^2/3$.
\begin{figure}[ht]
\centering
\includegraphics[scale=0.3]{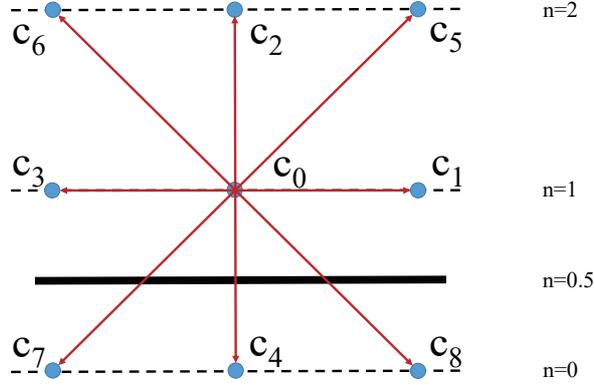}
\caption{ The half-way bounce back scheme in D2Q9 lattice model; the solid line denotes the bottom physical boundary while the dotted line denotes the computational grid.} \label{Fig1}
\end{figure}

By adopting a HBB scheme for Dirichlet boundary conditions~\cite{Ting2012General}, as shown in Fig.~\ref{Fig1}, Eq.~(\ref{eq48}) has a simple solution:
\begin{equation}\label{eq49}
  \phi_n=\phi_0 + \Delta \phi y_n(2-y_n)+\phi_s,
\end{equation}
where $y_n = (n-1/2)/N $ with N denoting the gride number in y direction, $\phi_s$ represents the $\emph{numerical slip }$ caused by the discrete effect of boundary condition \cite{He1997Analytic,Cui2016Discrete}.

We now turn to derive the \emph{numerical slip} by conducting some algebraic manipulations. From Eq.~(\ref{eq49}), we can get
\begin{equation}\label{eq50}
\phi_1=\phi_0+\frac{1}{2N}\left(2-\frac{1}{2N}\right)+\phi_s,
\end{equation}
\begin{equation}\label{eq51}
\phi_2=\phi_0+\frac{3}{2N}\left(2-\frac{3}{2N}\right)+\phi_s.
\end{equation}

With the help of the $u_x=0.1,u_y=0$, substituting Eq.~(\ref{eq46}) into Eq.~(\ref{eq45}) yields
\begin{equation}
\begin{aligned}\label{eq52}
 f^{n-1}_4+f^{n-1}_7+f^{n-1}_8=&\frac{2-k_1}{6}\phi_n+(k_1-1)\left[\frac{2-k_1}{6}\phi_{n-1}+(k_1-1)(f^{n-1}_4+f^{n-1}_7+f^{n-1}_8)+\frac{1}{6}\Delta tS-\left(1-\frac{k_1}{2}-\frac{k_2}{2}\right)\frac{2}{3k_2}\Delta tS\right]\\
 &+\frac{1}{6}\Delta tS-\left(1-\frac{k_1}{2}-\frac{k_2}{2}\right)\frac{2}{3k_2}\Delta tS
\end{aligned}
\end{equation}
rearranging above equation, we have
\begin{equation}\label{eq53}
k_1(f^{n-1}_4+f^{n-1}_7+f^{n-1}_8)=\frac{1}{6}\phi_n+\frac{k_1-1}{6}\phi_{n-1}+\frac{s_1}{2-k_1}\left[\frac{1}{6}\Delta tS-\left(1-\frac{k_1}{2}-\frac{k_2}{2}\right)\frac{2}{3k_2}\Delta tS\right]
\end{equation}

Based on HBB scheme, one can obtain
\begin{equation}\label{eq54}
f^1_2=-f^{1,+}_4+2w_4\phi_0,
\end{equation}
\begin{equation}\label{eq55}
f^1_5=-f^{1,+}_7+2w_7\phi_0,
\end{equation}
\begin{equation}\label{eq56}
f^1_6=-f^{1,+}_8+2w_8\phi_0.
\end{equation}

After a summation of Eqs.~(\ref{eq54})-(\ref{eq56}), we have
\begin{equation}\label{eq57}
 f^1_2+f^1_5+f^1_6=-( f^{1,+}_4+f^{1,+}_7+f^{1,+}_8)+\frac{1}{3}\phi_0.
\end{equation}

Substituting Eq.~(\ref{eq45}) into Eq.~(\ref{eq57}) results in the following equation
\begin{equation}\label{eq58}
f^1_2+f^1_5+f^1_6=-\left[\frac{2-k_1}{6}\phi_{1}+(k_1-1)(f^{1}_2+f^{1}_5+f^{1}_6)+\frac{1}{6}\Delta tS-\left(1-\frac{k_1}{2}-\frac{k_2}{2}\right)\frac{2}{3k_2}\Delta tS\right]+\frac{1}{3}\phi_0,
\end{equation}

If we substitute Eq.~(\ref{eq58}) into  Eq.~(\ref{eq43}), one can obtain
\begin{equation}\label{eq59}
 k_1(f^1_4+f^1_7+f^1_8)=\frac{2+k_1}{6}\phi_1-\frac{1}{3}\phi_0-\frac{2k_1}{3k_2}\Delta tS+\frac{1}{6}\Delta tS-\left(1-\frac{k_1}{2}-\frac{k_2}{2}\right)\frac{2}{3k_2}\Delta tS,
\end{equation}
then substituting Eq.~(\ref{eq53})into Eq.~(\ref{eq59}) yields
\begin{equation}\label{eq60}
3\phi_1-\phi_2-2\phi_0=\frac{4k_1}{k_2}\Delta tS+\frac{12(k_1-1)}{2-k_1}\left[\frac{1}{6}\Delta tS-\left(1-\frac{k_1}{2}-\frac{k_2}{2}\right)\frac{2}{3k_2}\Delta tS\right].
\end{equation}

Finally, the $\emph{numerical slip}$ can be obtain through substituting Eq. (\ref{eq50}),~Eq. (\ref{eq51}) into Eq. (\ref{eq60}),

\begin{equation}\label{eq61}
\phi_s=\frac{3k_1k_2-8k_1-12k_2+16}{12k_1k_2}\frac{\Delta \phi}{N^2}.
\end{equation}

Let $\phi_s=0$, we can obtain the following expression,
\begin{equation}\label{eq62}
k_2=\frac{8(k_1-2)}{3(k_1-4)},
\end{equation}
which can be used to eliminate the $\emph{numerical slip}$.

\section{Numerical results and discussion}

To test the stability and accuracy of present model, some simulations of isotropic CDEs and anisotropic CDEs are performed. The HBB scheme \cite{Ting2012General} is adopted to treat the Dirichlet boundary conditions. In our simulations, the following global relative error (GRE) is used to measure the accuracy of the present B-TriRT model,
\begin{equation}\label{eq63}
  GRE=\frac{\sum_{i,j} |\phi(\textbf{x},t)-\phi^{*}(\textbf{x},t)|}{\sum_{i,j} |\phi^{*}(\textbf{x},t)|},
\end{equation}
where $\phi$ and $\phi^{*}$ are the numerical and analytical solution, respectively. Besides, for the steady flows, the following convergent criterion is adopted,
\begin{equation}\label{eq64}
  \frac{\sum_{i,j} |\phi(\textbf{x},t+1000\Delta t)-\phi(\textbf{x},t)|}{\sum_{i,j} |\phi(\textbf{x},t+1000\Delta t)|}<10^{-10}.
\end{equation}
Unless otherwise stated, the distribution function $f_i(\textbf{x},t)$ is initialized by its equilibrium distribution function $f_i^{eq}(\textbf{x},t)$, and the tunable parameter $Z$ in MLBM is set as 0.0001 for a satisfactory accuracy.

\subsection{Isotropic CDEs}
\subsubsection{A steady diffusion problem}
In this part, we will validate the expression [Eq.~(\ref{eq62})] by conducting some simulations of a steady diffusion problem in the physical region $[0, L]\times [0, L]$, which can be described by Eqs.~(\ref{eq31})-(\ref{eq32}).

\begin{figure}[ht]
\centering
\subfigure[]{ \label{fig2:a}
\includegraphics[scale=0.4]{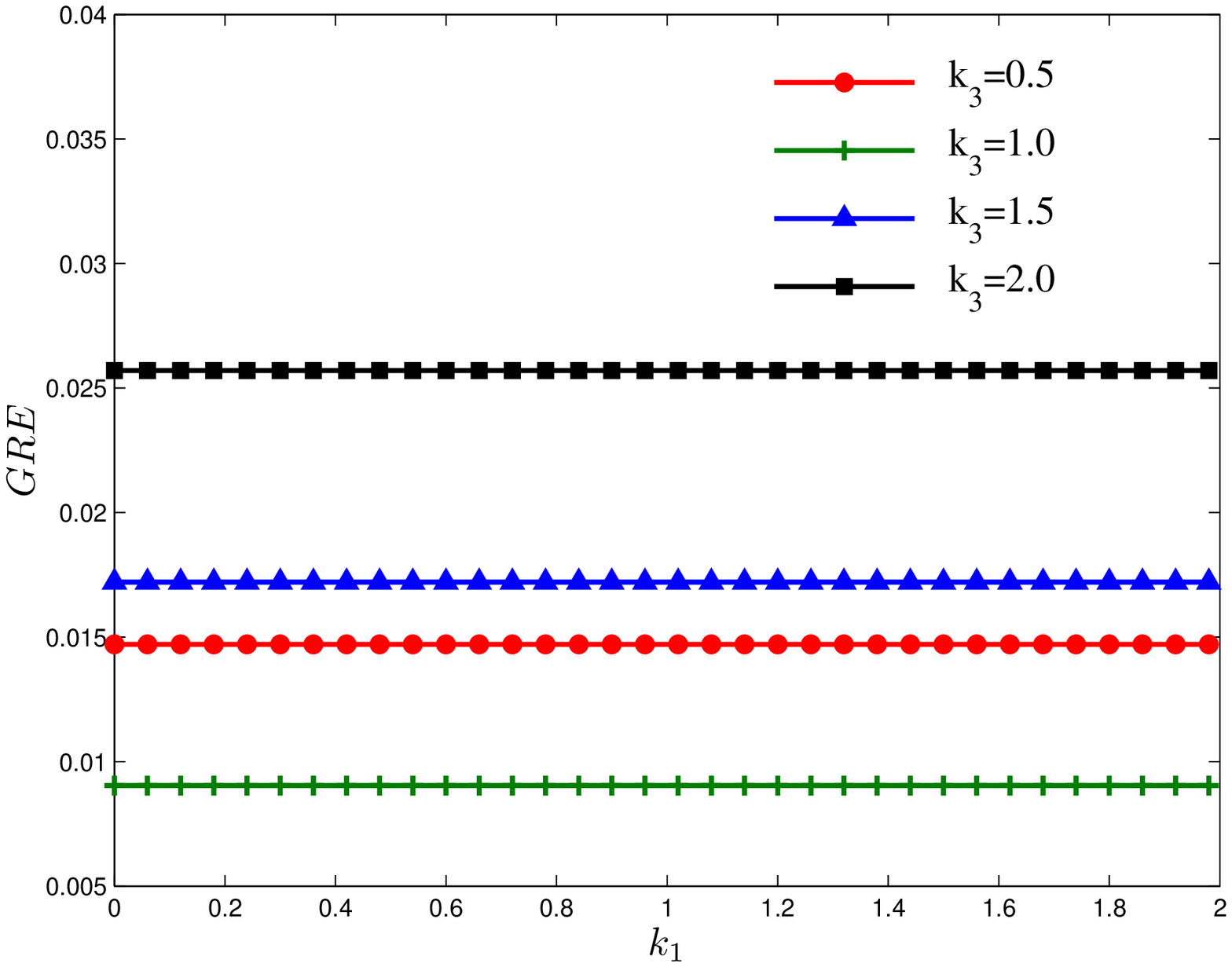}}
\subfigure[]{ \label{fig2:b}
\includegraphics[scale=0.4]{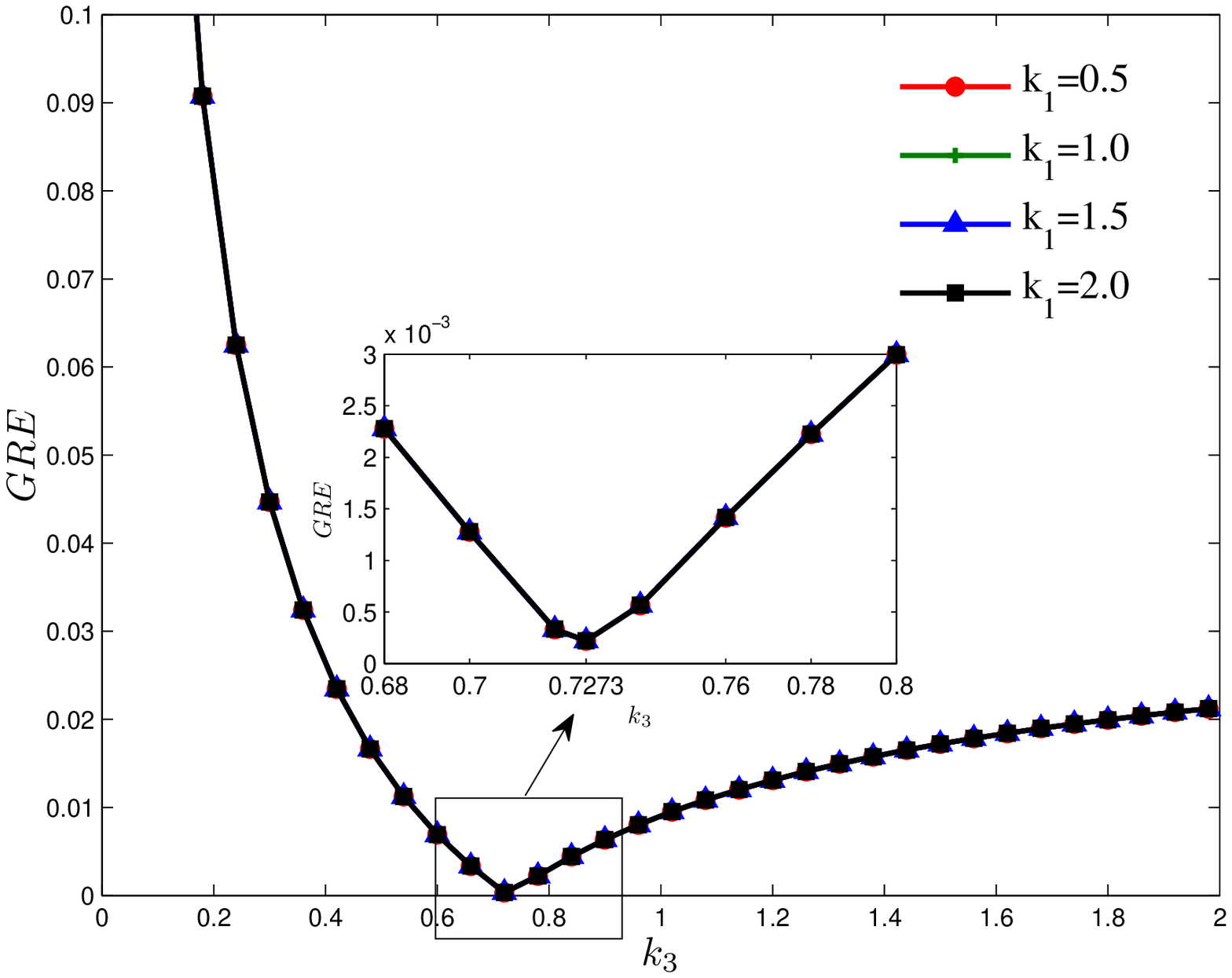}}
\caption{The global relative errors at different relaxation parameters [(a): the effects of $k_0$  (b): the effects of $k_2$].  } \label{Fig2}
\end{figure}

For isotropic CDEs, the relaxation parameters could be simplified as $\textbf{K}_1=k_1\textbf{I},~\textbf{K}_2=k_2\hat{\textbf{I}}$. In the simulations, $L=1.0,~u_x=0.1,~\alpha=0.1,~\phi_L=1.0,~\phi_0=0,~\Delta x = L/5,~k_1=1.25$. We first investigated the effects of tunable relaxation parameters $k_0$ and $k_2$ on the accuracy of present B-TriRT model, and presented the results in Fig.~\ref{Fig2}. From Fig.~\ref{fig2:a}, one can observe that the relaxation parameter $k_0$  has no distinct effect on the accuracy of present model. The reason is simple: from the matrix analysis, one can find that the parameter $k_0$ is related to the zero-order moment of non-equilibrium distribution function $f_i^{neq}$, and has no influence on the derivation of Eq.~(\ref{eq48}). On the other hand, the relaxation parameter $k_2$ plays an important role in the accuracy of present model. From the inserted figure in Fig.~\ref{fig2:b}, the optimal $k_2$ for this specific simulation is 0.7273, which is consistent with the relational expression $k_2= [8(k_1-2)]/[3(k_1-4)]=0.7273.$

\begin{figure}[ht]
\centering
\includegraphics[scale=0.5]{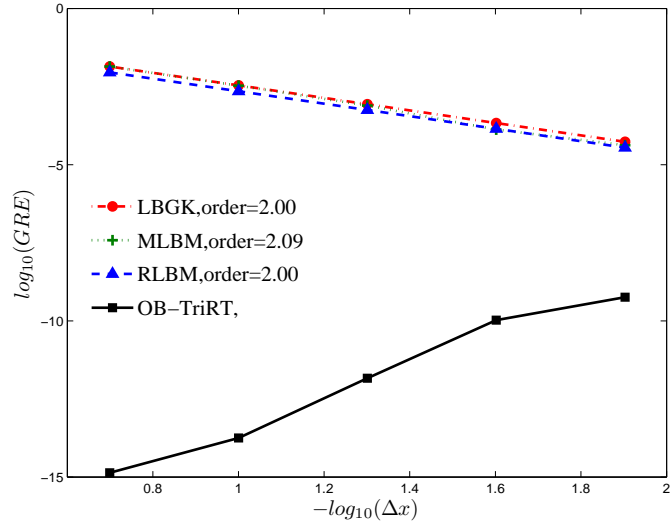}
\caption{ The global relative errors of different models at different mesh sizes.} \label{Fig3}
\end{figure}

\begin{figure}[ht]
\centering
\includegraphics[scale=0.5]{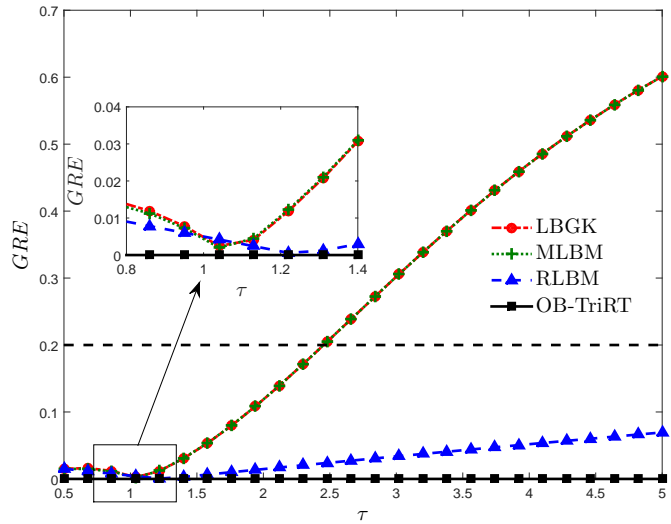}
\caption{ The global relative of different models errors at different relaxation times.} \label{Fig4}
\end{figure}

Furthermore, to test the convergence rate of present model, we also perform some simulation with different mesh  resolutions ($\Delta x=L/5,~L/10,~L/20,~L/40,~L/80$), and presented the $GREs$ in Fig.~\ref{Fig3}, where the optimal block triple-relaxation-time (OB-TriRT) lattice Boltzmann model corresponds to the case with $k_0 = 1,~k_1 = 1/\tau,~k_2 = [8(k_1-2)]/[3(k_1-4)].$ As shown in Fig.~\ref{Fig3}, both LBGK, MLBM and RLBM have a second-order accuracy in space, while the magnitude of $GREs$ in OB-TriRT model is around $10^{-10}$, which is close to the machine Error. Actually, the $GREs$ of OB-TriRT model could be zero since the \emph{numerical slip} is completely eliminated, and the present $GREs$ are caused by the machine error.

Then, we carried out some simulations under different value of relaxation time $\tau$ to investigate the stability of different models. From Fig.~\ref{Fig4}, one can clearly find that the OB-TriRT model is more stable than other models. Besides, the RLBM is more stable than LBGK and MLBM, the result is consistent with the  previous work~\cite{Wang2015Regularized}. Moreover, the results in Fig.~\ref{Fig4} also indicate that the \emph{numerical slip}  is completely eliminated by the adoption of expression $k_2=[8(k_1-2)]/[3(k_1-4)].$

{\color{blue}Finally, we conducted a simple  test on an Intel Core  i5-8250U CPU and gave a comparison in the  Tab.~\ref{tab2}. From the table, one can obtain that the present model doesn't show the best numerical efficiency due to the calculation of first-order and second-order moments of the non-equilibrium distribution. However, the present model has better numerical efficiency than the MRT model.}

\begin{table}
\centering
\caption{{\color{blue}The comparison of numerical efficiency among different models  }}\label{tab2}
\begin{tabular}{c|c|c|c|c}
 \hline\hline
  \multicolumn{1}{c|}{ \multirow{2}*{Model} }&\multicolumn{2}{c|}{Mesh: $\Delta x=1/5$}&\multicolumn{2}{c}{Mesh: $\Delta x=1/10$}\\
\cline{2-5}
\multicolumn{1}{c|}{}& Time of 100 steps (s) & Time of steady state (s) & Time of 100 steps (s) & Time of steady state (s) \\
\hline
  LBGK &0.005 & 0.022 & 0.017& 0.276 \\
  MLBM & 0.020& 0.088  &0.062 &1.214 \\
  RLBM &0.020& 0.088&0.062&1.201  \\
  MRT&0.051  &0.264 & 0.162&3.010 \\
 OB-TriRT  &0.039&0.231&0.149&2.786\\
  \hline\hline
\end{tabular}
\end{table}

\subsubsection{Convection-diffusion equation with a constant velocity}
Now, we consider the following linear isotropic convection-diffusion equation,
\begin{equation}\label{eq65}
 \partial_t \phi + \nabla \cdot (\phi \textbf{u}) = \nabla \cdot (\alpha\nabla \phi)+S, \\
\end{equation}
where $\textbf{u}=(u_x,u_y)^T$ is a constant velocity, $\alpha$ is the diffusion coefficient. $S$ is the source term, which is given by
$S=\textrm{exp}[(1-2\pi^2\alpha)t]{\textrm{sin}[\pi(x+y)]+\pi(u_x+u_y)\textrm{cos}[\pi(x+y)]}$. Under the proper initial and boundary conditions, the analytical solution of this problem can be given as
\begin{equation}\label{eq66}
  \phi(x,y,t) = \textrm{exp}[(1-2\pi\alpha)t]\textrm{sin}[\pi(x+y)]].
\end{equation}

\begin{figure}[ht]
\centering
\subfigure[]{ \label{fig5:a}
\includegraphics[scale=0.4]{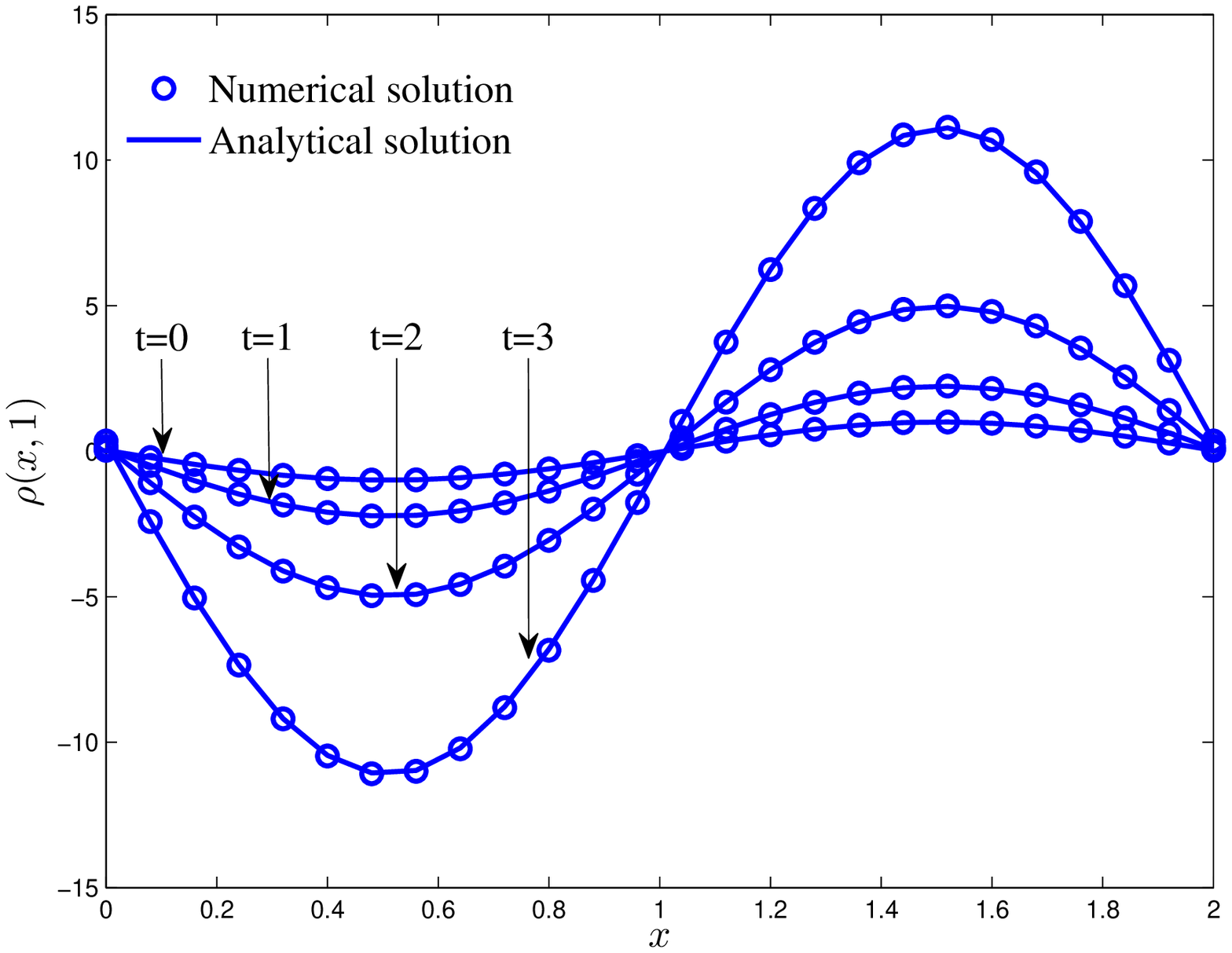}}
\subfigure[]{ \label{fig5:b}
\includegraphics[scale=0.4]{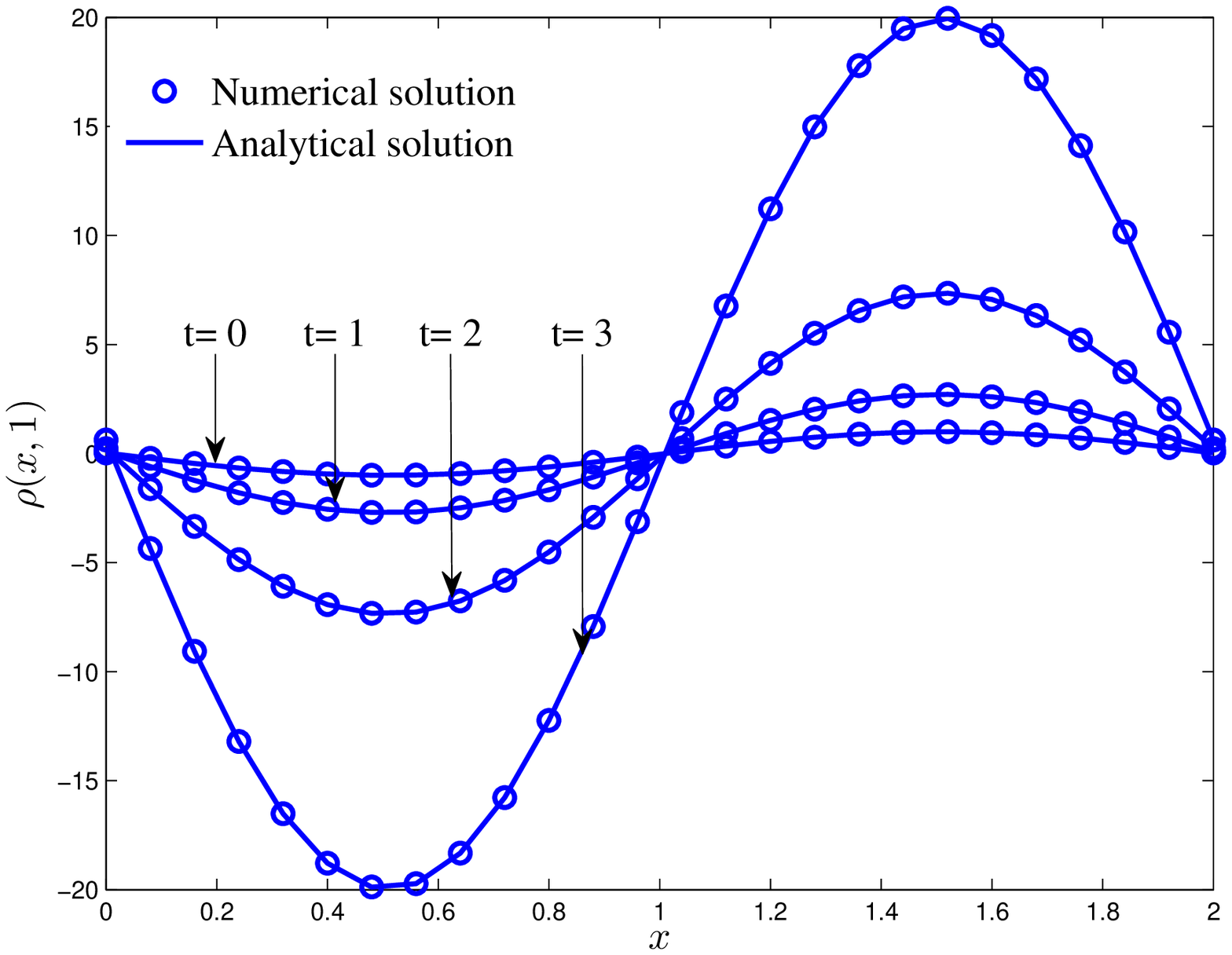}}
\caption{Numerical and analytical solutions at different time and diffusion coefficients [(a): $\alpha=0.01$   (b): $\alpha=0.0001$].} \label{Fig5}
\end{figure}

In the simulations, the computational domain is fixed to be $[0,2]\times[0,2]$. We first performed some simulations under different time and different diffusion coefficients, and presented the results in Fig.~\ref{Fig5} where $c=1.0,~u_x = u_y = 0.1$, the mesh size is $200\times 200$. As we can see from the figure, the numerical solutions agree well with analytical solutions, even with a small diffusion coefficient. The $GREs$ at time $t=3$ are $3.009\times 10^{-3}$ for $\alpha=0.01$ and $1.745\times 10^{-4}$ for $\alpha=0.0001$.

\begin{figure}[ht]
\centering
\includegraphics[scale=0.5]{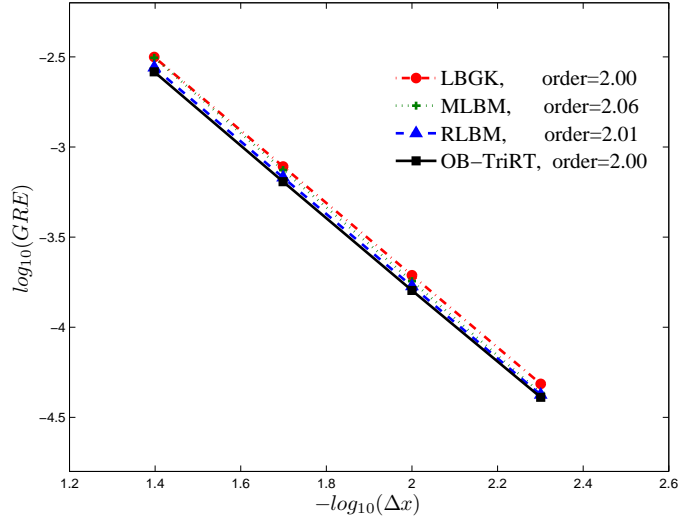}
\caption{ The global relative errors of different models at different mesh sizes.} \label{Fig6}
\end{figure}

We also tested the convergence rate of present model, and conducted a number of simulations under different mesh resolutions $\Delta x = 1/25,~1/50,~1/100,~1/200$ with $\tau$ being fixed at 0.8. The results shown in Fig.~\ref{Fig6} indicate that all the LBGK, MLBM, RLBM and OB-TriRT model have a second-order accuracy in space, and the OB-TriRT model is more accurate than other models.  In addition, the $GREs$ of OB-TriRT model are different from the results in Fig.~\ref{Fig3}, it means that the \emph{numerical slip}  hasn't been completely eliminated. However, the adoption of expression $[k3= 8(k_1-2)]/[3(k_1-4)]$ can still bring some benefits in accuracy.

Furthermore, to investigate the stability of present OB-TriRT model for the problem with different convection velocity ($u_0=0.01,~0.1,~1.0,~2.5$), we carried out some simulations with $\Delta x=1/50,~c=5.0,~\tau=0.8,~\alpha=0.01,~t=1,~u_x=u_y=u_0$, and presented a quantitative comparison in Tab.~\ref{tab3}. As shown in this table, all models are unstable for the case $u_0=2.5$ except for the OB-TriRT model, which indicates that the present OB-TriRT model is  more accurate than other models. Accordingly, the results also indicate that the expression $[k3= 8(k_1-2)]/[3(k_1-4)]$ could be used to improve the stability of present model.
\begin{table}[ht]

\centering
\caption{The global relative errors of different models at different convection velocities ("-":unstable)}\label{tab3}
\begin{tabular}{c|c|c|c|c}
  \hline \hline
  Model& $u_0=0.01$ & $u_0=0.1$ & $u_0=1.0$ & $u_0=2.5$ \\
    \hline
  LBGK & 7.5225 $\times~10^{-4}$ & 7.7943 $\times~10^{-4}$ & 2.6979 $\times~10^{-3}$& - \\
  MLBM & 3.9366 $\times~10^{-4}$ & 4.5053 $\times~10^{-4}$ & 2.6503 $\times~10^{-3}$ & -\\
  RLBM & 6.9176 $\times~10^{-4}$& 6.7784 $\times~10^{-4} $& 3.4796 $\times~10^{-3} $& - \\
  OB-TriRT & 5.8108 $\times~10^{-4}$ & 6.5226 $\times~10^{-4}$ & 2.1399 $\times~10^{-3}$& 1.7531 $\times~10^{-2}$\\
  \hline\hline

\end{tabular}
\end{table}

\subsubsection{Burgers-Fisher equation}
The Burgers-Fisher equation (BFE), as a special case of the nonlinear CDE and a popular benchmark problem, is also adopted to test the present OB-TriRT model. The BFE can be expressed as \cite{Wazwaz2005The}

\begin{equation}\label{eq67}
  \partial_t \phi+ a \phi^{\delta}\partial_x\phi = \alpha(\partial_{xx}\phi+\partial_{yy}\phi)+b\phi(1-\phi^{\delta}),
\end{equation}
where $\delta,~b$ are constants, $a, \alpha $ are the constant convection and diffusion coefficients. Under the proper initial and boundary conditions, the analytical solution can be expressed as
\begin{equation}\label{eq68}
 \phi(x,y,t)=\left\{\frac{1}{2}+\frac{1}{2}\textrm{tanh}[A(x+y-\omega t)]\right\}^{\frac{1}{\delta}},
\end{equation}
where A and $\omega$ are defined as
\begin{equation}\label{eq69}
  A=-\frac{a\delta}{4\alpha(\delta+1)},~~ \omega=\frac{a^2+2b\alpha(\delta+1)}{a(\delta+1)}.
\end{equation}

\begin{figure}[ht]
\centering
\subfigure[]{ \label{fig7:a}
\includegraphics[scale=0.4]{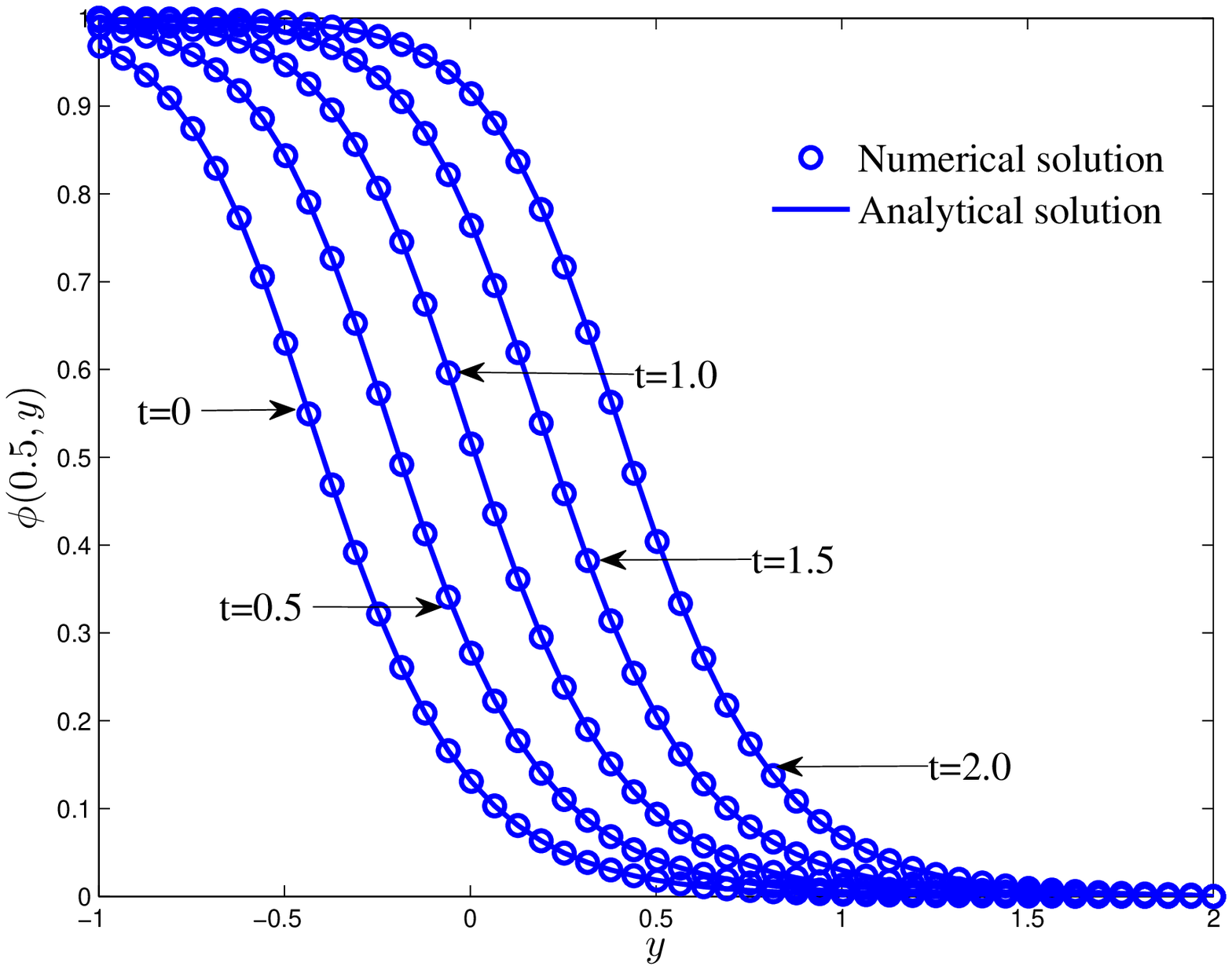}}
\subfigure[]{ \label{fig7:b}
\includegraphics[scale=0.4]{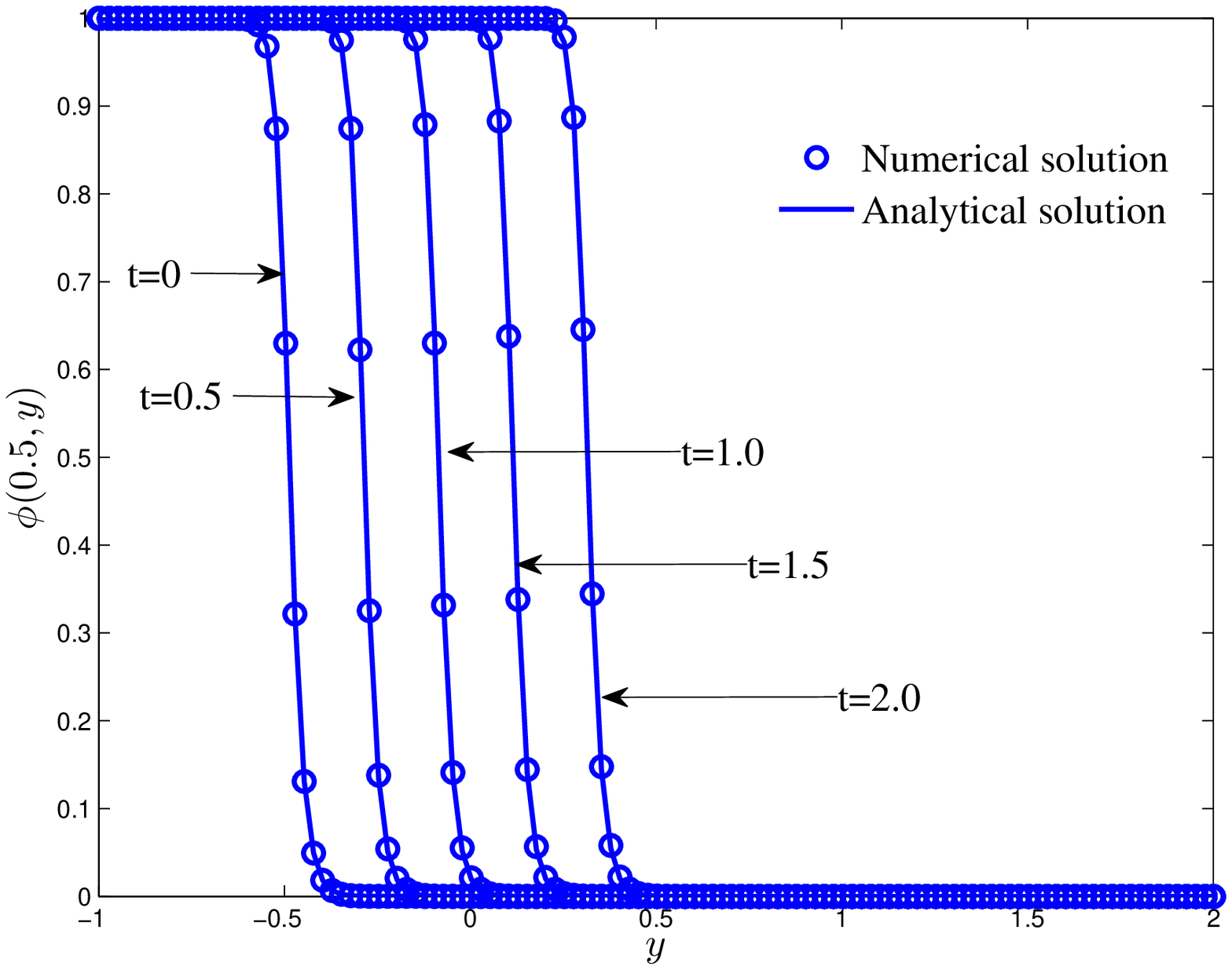}}
\caption{Numerical and analytical solutions at different time and diffusion coefficients [(a): $\alpha=0.05$   (b): $\alpha=0.0005$].} \label{Fig7}
\end{figure}

The simulations are performed on domain $[-1,2]\times[-1,2]$, and the effect of coefficients $a$ and $\alpha$ are mainly investigated for different models. Firstly, we tested the capacity of present OB-TriRT model in solving the nonlinear BFE, and presented the numerical and analytical solutions at different time and different diffusion coefficients in Fig.~\ref{Fig7} where $\Delta x=1/160,~c=5,~a=2.0,~b=0.05,~\delta=1.5$. As shown in the figure, the numerical solutions are in good agreement with analytical solutions. We also calculated the difference between the analytical solutions and numerical solutions at time $t=2.0$, and found that $GREs=4.7266\times10^{-5},~1.1247\times10^{-4} $ for $\alpha=0.05, 0.005$ respectively. The results show that the present OB-TriRT model has the ability in solving nonlinear CDEs, even with a small diffusion coefficient.

\begin{figure}[ht]
\centering
\includegraphics[scale=0.5]{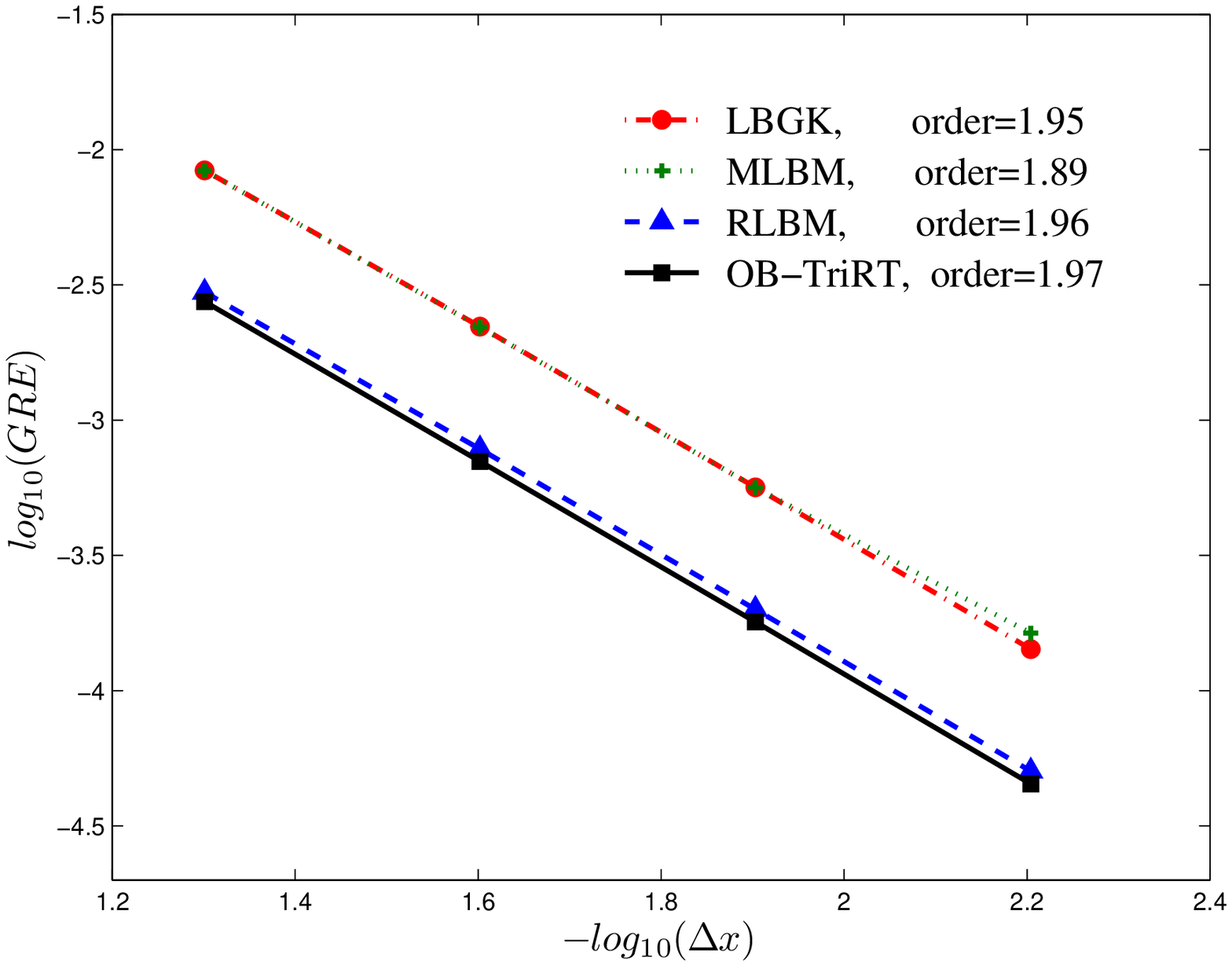}
\caption{ The global relative errors of different models at different mesh sizes.} \label{Fig8}
\end{figure}

Then, we tested the convergence rate of present OB-TriRT model by conducting a number of simulations under different mesh resolutions ($\Delta x= 1/20,~1/40,~1/80,~1/160$) with $a=2.0, ~b=0.05,~\delta=1.5,~\alpha=0.01, t=1.0.$ The results shown in Fig.~\ref{Fig8} indicate that present OB-TriRT model has a second-order convergence rate in space. In addition, the $GREs$ of OB-TriRT model are smaller than other models, which implies that the expression $[k3= 8(k_1-2)]/[3(k_1-4)]$ can be used to improve the accuracy of present model.

\begin{table}[ht]
\centering
\caption{The global relative errors of different models at different convection coefficients with $\alpha=0.05$ ("-":unstable)}\label{tab4}
\begin{tabular}{c|c|c|c|c}
  \hline \hline
  Model& $a=1$ & $a=2$ & $a=3$ & $a=4$ \\
    \hline
  LBGK & 1.1958 $\times ~10^{-3}$ & 3.2058 $\times~10^{-3}$ & 4.1915 $\times~10^{-3}$& - \\
  MLBM & 1.2038 $\times ~10^{-3}$ & 3.2015 $\times~10^{-3}$ & 1.1927 $\times~10^{-3}$ & -\\
  RLBM & 5.2339 $\times~10^{-4}$ & 9.9594 $\times~10^{-4} $& 1.1328 $\times~10^{-3}$ & - \\
  OB-TriRT & 5.0193 $\times~10^{-4}$ & 9.5559 $\times~10^{-4}$ & 1.0653 $\times~10^{-3}$ & 2.2140 $\times~10^{-3}$\\
  \hline\hline

\end{tabular}
\end{table}

\begin{table}[ht]
\centering
\caption{The global relative errors of different models at different convection coefficients with $a=2$ ("-":unstable)}\label{tab5}
\begin{tabular}{c|c|c|c|c}
  \hline \hline
  Model& $\alpha=0.005$ & $\alpha=0.05$ & $\alpha=0.1$ & $\alpha=0.5$ \\
    \hline
  LBGK & -  & 3.2058 $\times~ 10^{-3}$ & 5.7829 $\times~ 10^{-3}$& - \\
  MLBM & - & 3.2015 $\times ~10^{-4}$ & 5.7852$\times ~10^{-3}$ & -\\
  RLBM & 4.7209 $\times ~10^{-3}$ & 9.9594 $\times ~10^{-4} $& 1.8174$\times~ 10^{-3} $& 1.7866 $\times~ 10^{-3}$\\
  OB-TriRT & 9.6078 $\times ~10^{-3}$ & 9.5559 $\times ~10^{-4}$ & 2.0497$\times~ 10^{-3}$& 2.0271 $\times~ 10^{-3}$\\
  \hline\hline

\end{tabular}
\end{table}
Finally, we turn to investigate the effects of coefficients $a$ and $\alpha$, and present a comparison of different models in Tab.~\ref{tab4} and Tab.~\ref{tab5} where $\Delta x=1/40,~c=5,~b=0.05, ~\delta=1.5$. As shown in Tab.~\ref{tab4}, when $a$ increases from 1 to 4, the $GREs$ increase. In addition, when $a$ is increased to 4, all models are unstable apart from OB-TriRT model. Moreover, the results show that the OB-TriRT model could be more accurate than other models under the  present parameters. From Tab.~\ref{tab5}, one can obtain that the LBGK and MLBM are unstable at $\alpha=0.005$ and $\alpha=0.5$. Moreover, under the  present parameters, the $GREs$ of OB-TriRT model are not always less than RLBM, that implies the expression $k_2=[8(k_1-2)]/ [3(k_1-4)]$ may be not the optimal choice for some nonlinear CDEs. However, as discussed above, the adoption of  this expression still could give a satisfactory results for the complicated nonlinear problems.

\subsection{Anisotropic CDEs}
\subsubsection{Gaussian hill problem}

In the following parts, we turn to test the capacity of present B-TriRT model in solving anisotropic CDEs. On the one hand, the anisotropic problems could not be solved directly by the previous LBGK model \cite{Baochang2009Lattice} due to the diffusion coefficient is not a scalar variable. On the other hand, although the MRT model has the ability in the study of the anisotropic problems, it is not convenient to determine the free relaxation parameters, and the additional  multiple-relaxation collision will reduce the computational efficiency.  It should be mentioned that, for anisotropic CDEs, the parameter related to diffusion coefficient is a matrix $\textbf{K}_1$ rather than a constant $k_1$, which can be determined by Eq.~(\ref{eq30}). Moreover, the expression Eq.~(\ref{eq62}) between $k_1$ and $k_2$ no longer exists. Without loss of generality, the parameter $k_2$ is set to be 1 for all simulations of the anisotropic CDEs. Firstly, we consider a classic benchmark example named Gaussian hill, which can be described by

\begin{equation}\label{eq70}
  \partial_t \phi+ \nabla \cdot(\phi\textbf{u})=\nabla \cdot (\textbf{A}\cdot \nabla \phi),
\end{equation}
where $\textbf{u}=(u_x,u_y)^T$ is a constant velocity, $\textbf{A}$ is a constant diffusion tensor. The analytical solution to this problem can be given by

\begin{equation}\label{eq71}
\phi(\textbf{x},t)=\frac{\phi_0}{2\pi\sqrt{|det(\sigma_t)|}}\textrm{exp}\left\{-\frac{1}{2}\sigma_t^{-1}:[(\textbf{x}-\textbf{u}t)(\textbf{x}-\textbf{u}t)]\right\},
\end{equation}
where $\sigma_t=\sigma_0^2\textbf{I}+2\textbf{A}t,|det(\sigma_t)|$ is absolute value of the determinant of $\sigma_t$, $\sigma_t^{-1}$ is the inverse matrix of $\sigma_t$. In our simulations, the computational domain is fixed on $[-1,1]\times[-1,1],~\sigma_0=0.01,~u_x=u_y=0.01$, and the following diffusion matrices are considered,

\begin{equation}\label{eq72}
          \textbf{A}=\left[\left(
               \begin{array}{cc}
                1&0\\
                0&1
               \end{array}
             \right),\quad
             \left(
               \begin{array}{cc}
                1&0\\
                0&2
               \end{array}
             \right),\quad
             \left(
               \begin{array}{cc}
                1&1\\
                1&2
               \end{array}
             \right)\right] \times 10^{-3},
\end{equation}
which are usually denoted as isotropic, diagonally anisotropic and fully anisotropic diffusion problems \cite{chai2016multiple}.

\begin{figure}[ht]
\centering
\subfigure[]{ \label{fig9:a}
\includegraphics[scale=0.4]{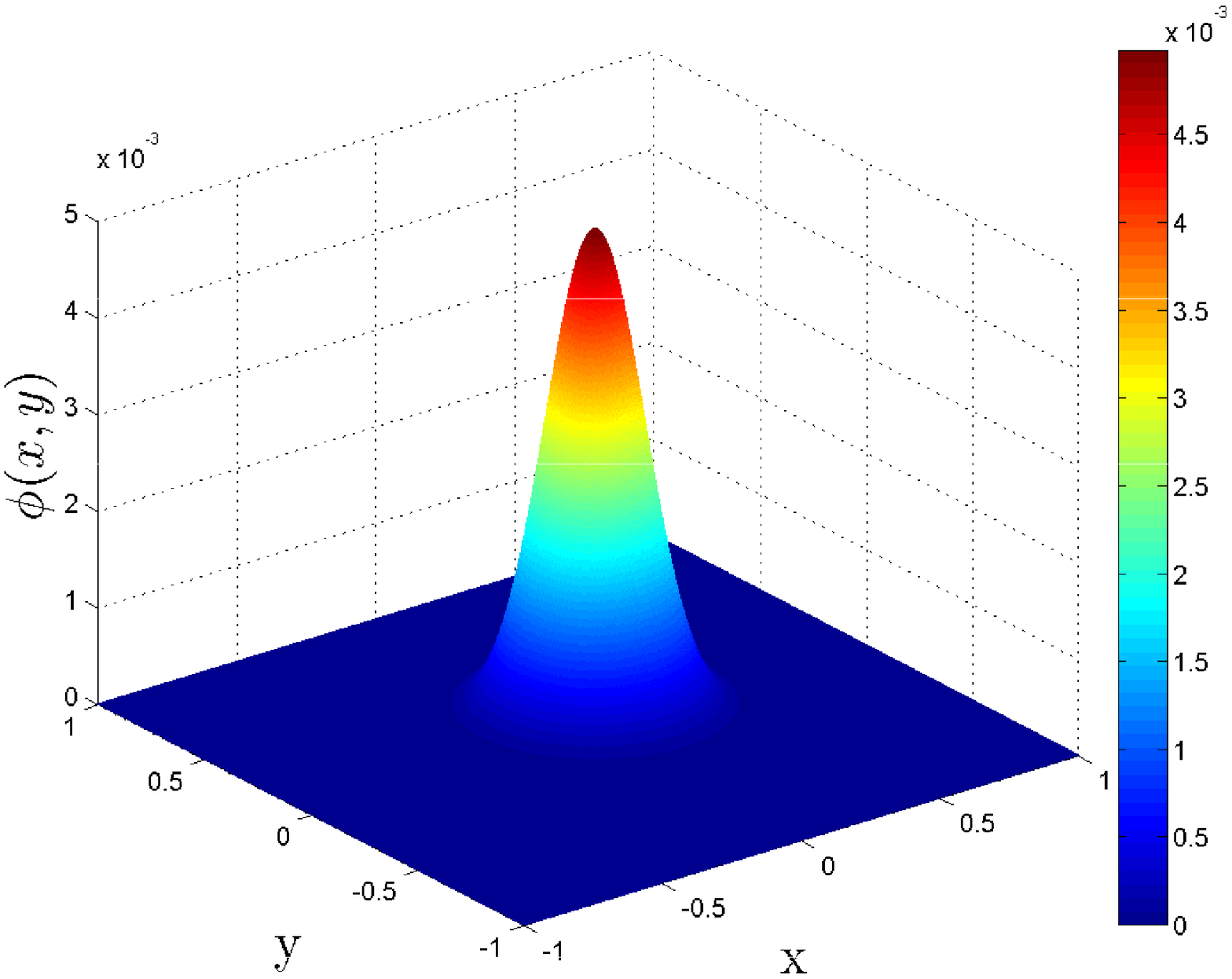}}
\subfigure[]{ \label{fig9:b}
\includegraphics[scale=0.4]{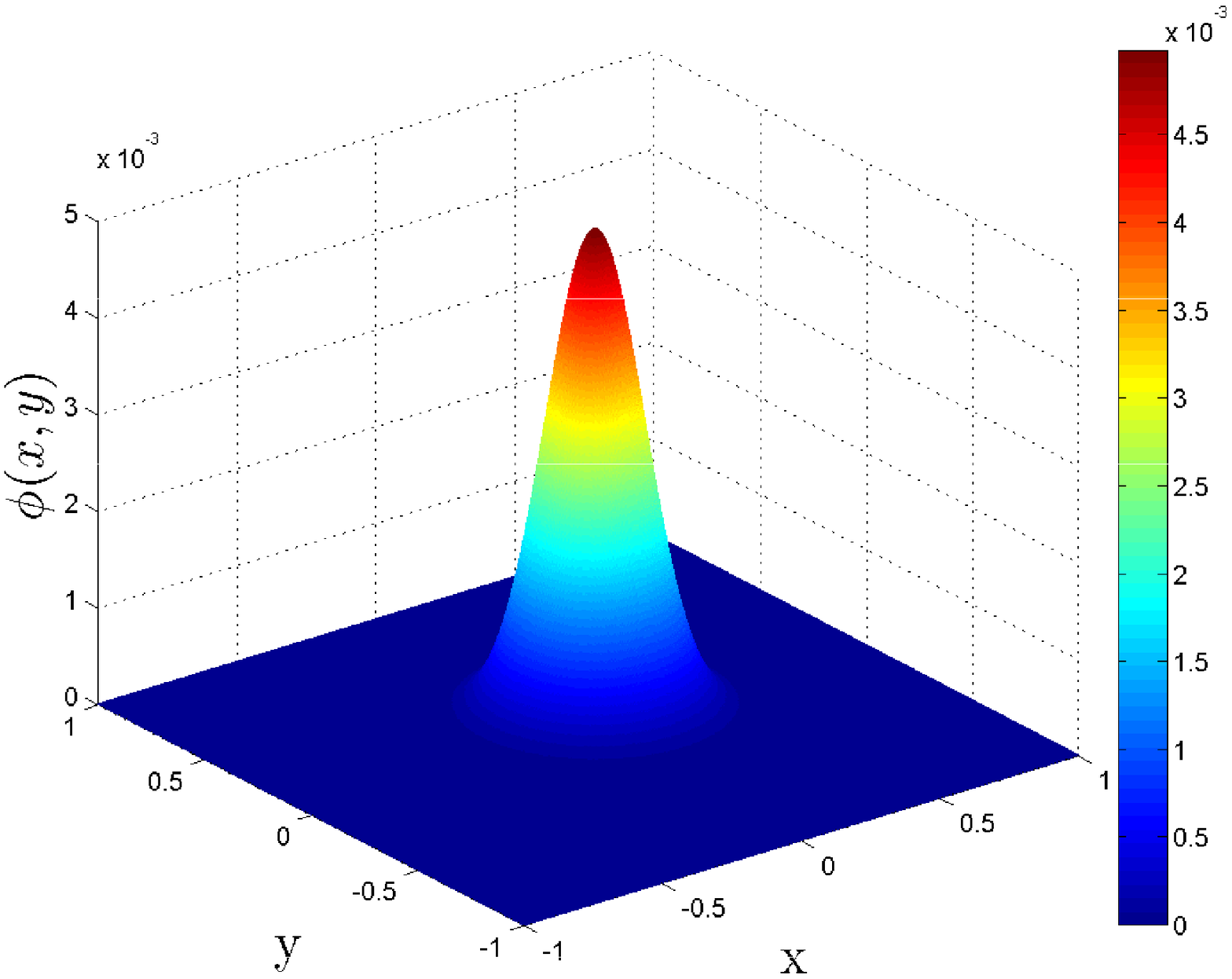}}
\caption{Numerical and analytical solutions at $t=10$  [(a): numerical solution   (b): analytical solution].} \label{Fig9}
\end{figure}

\begin{figure}[ht]
\centering
\subfigure[]{ \label{fig10:a}
\includegraphics[scale=0.4]{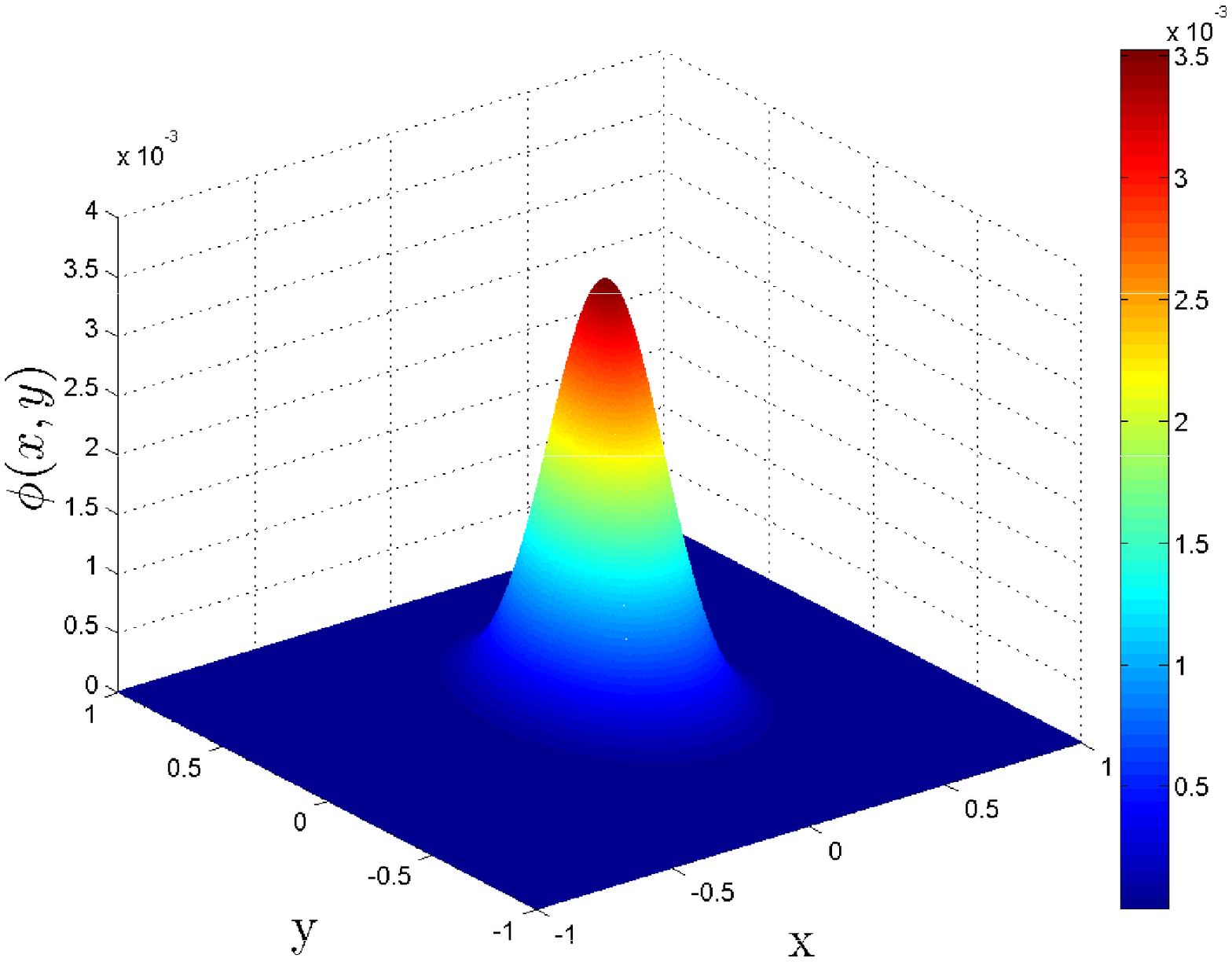}}
\subfigure[]{ \label{fig10:b}
\includegraphics[scale=0.4]{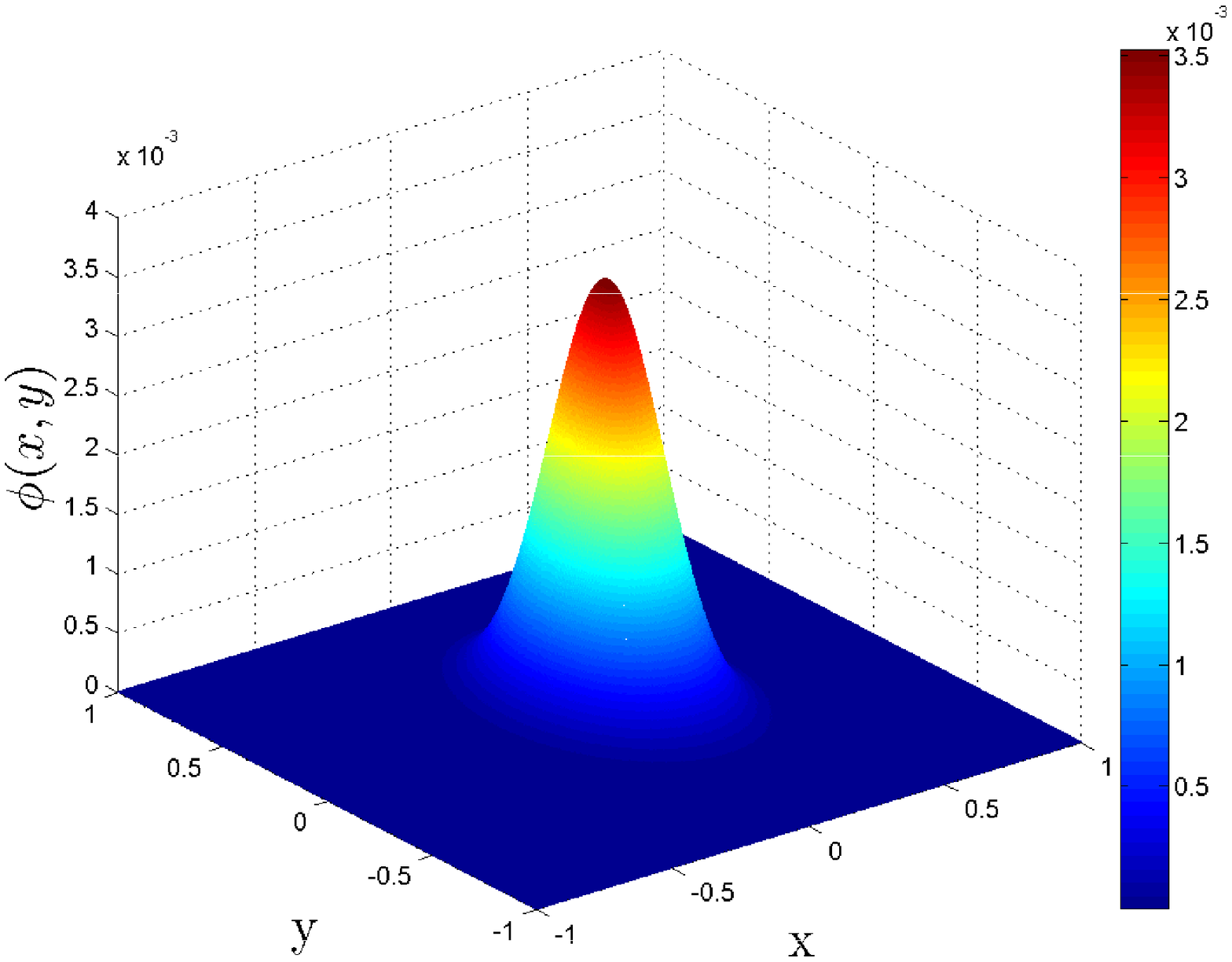}}
\caption{Numerical and analytical solutions at $t=10$  [(a): numerical solution   (b): analytical solution].} \label{Fig10}
\end{figure}

\begin{figure}[ht]
\centering
\subfigure[]{ \label{fig11:a}
\includegraphics[scale=0.4]{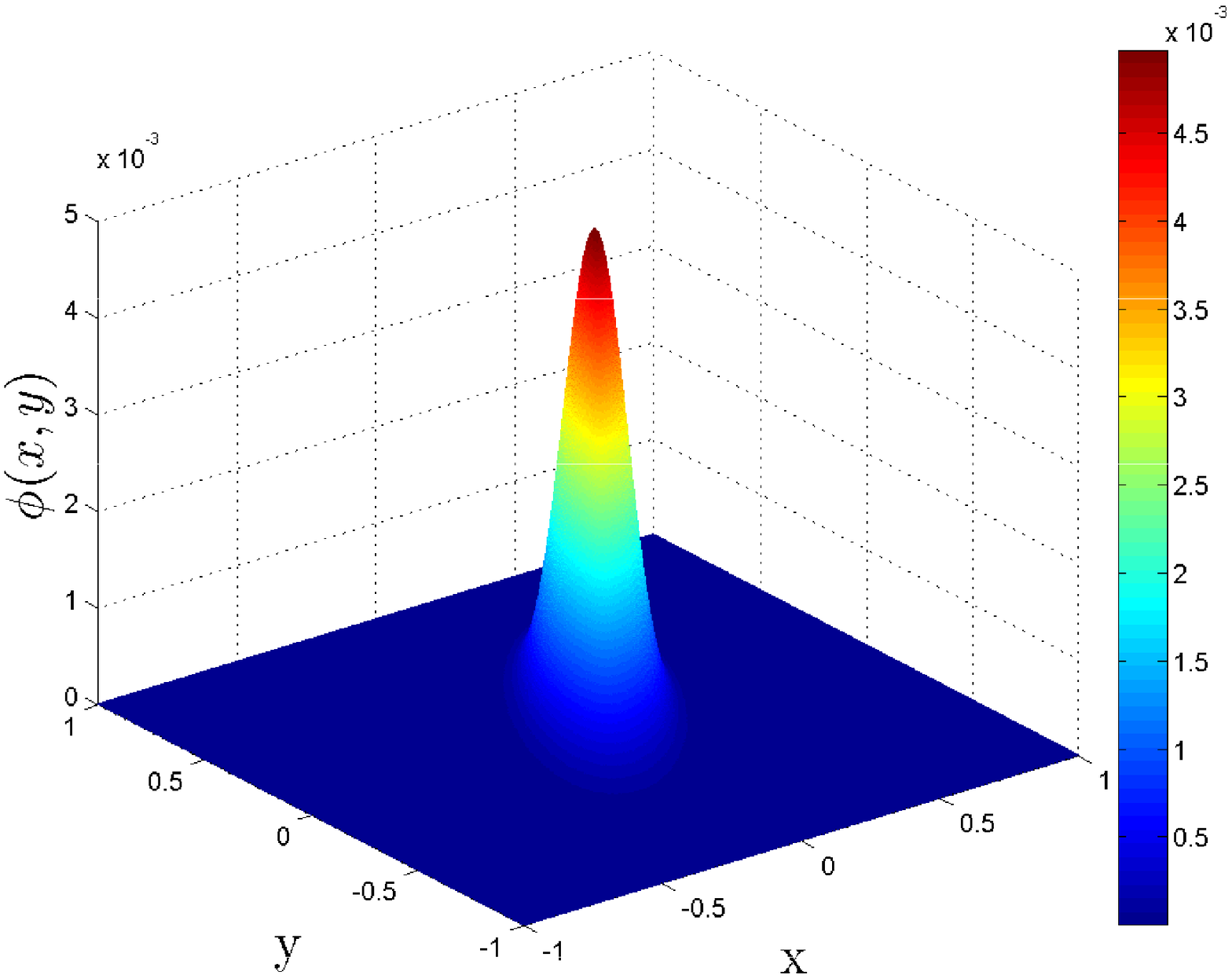}}
\subfigure[]{ \label{fig11:b}
\includegraphics[scale=0.4]{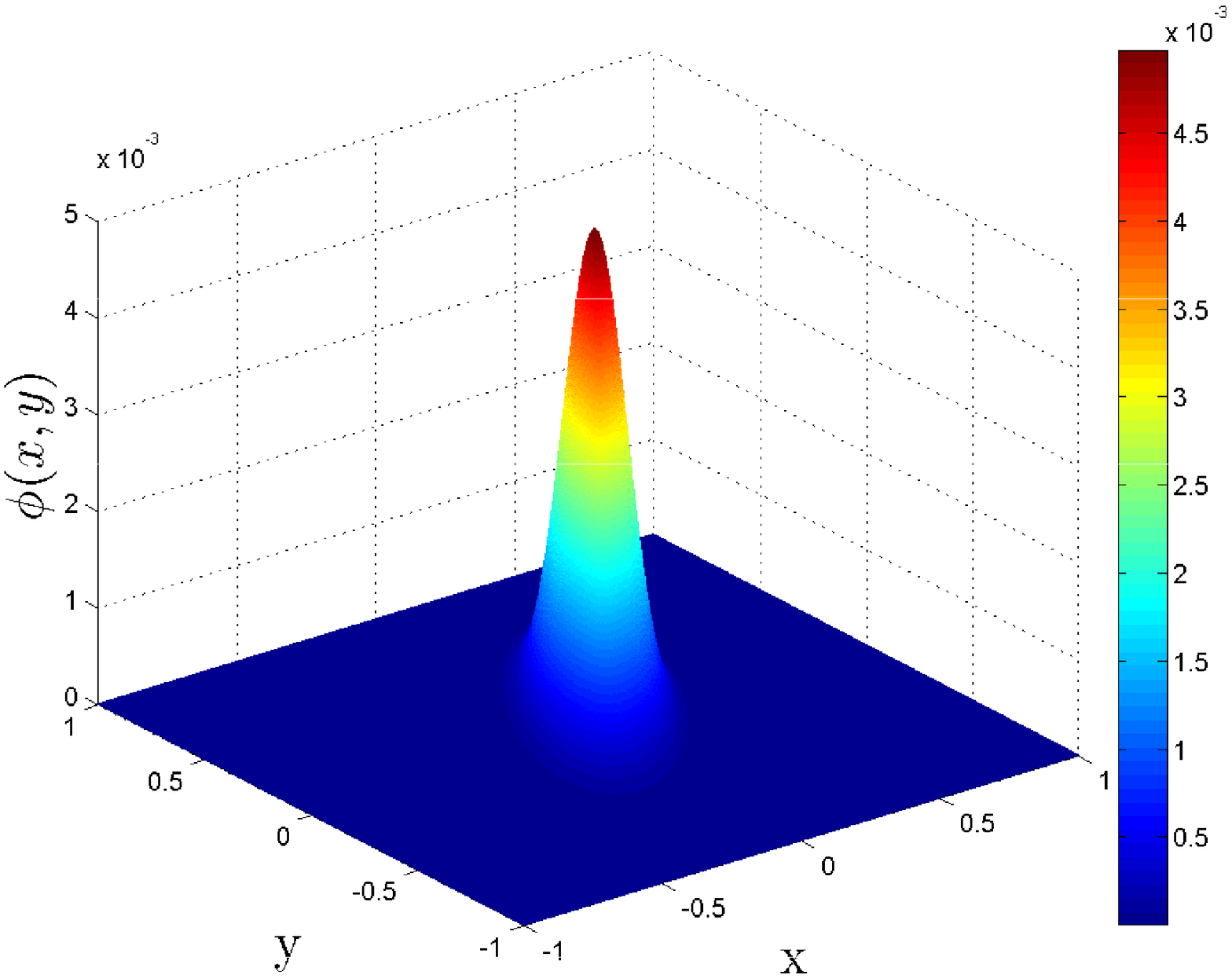}}
\caption{Numerical and analytical solutions at $t=10$ [(a): numerical solution   (b): analytical solution].} \label{Fig11}
\end{figure}

\begin{figure}[ht]
\centering
\includegraphics[scale=0.5]{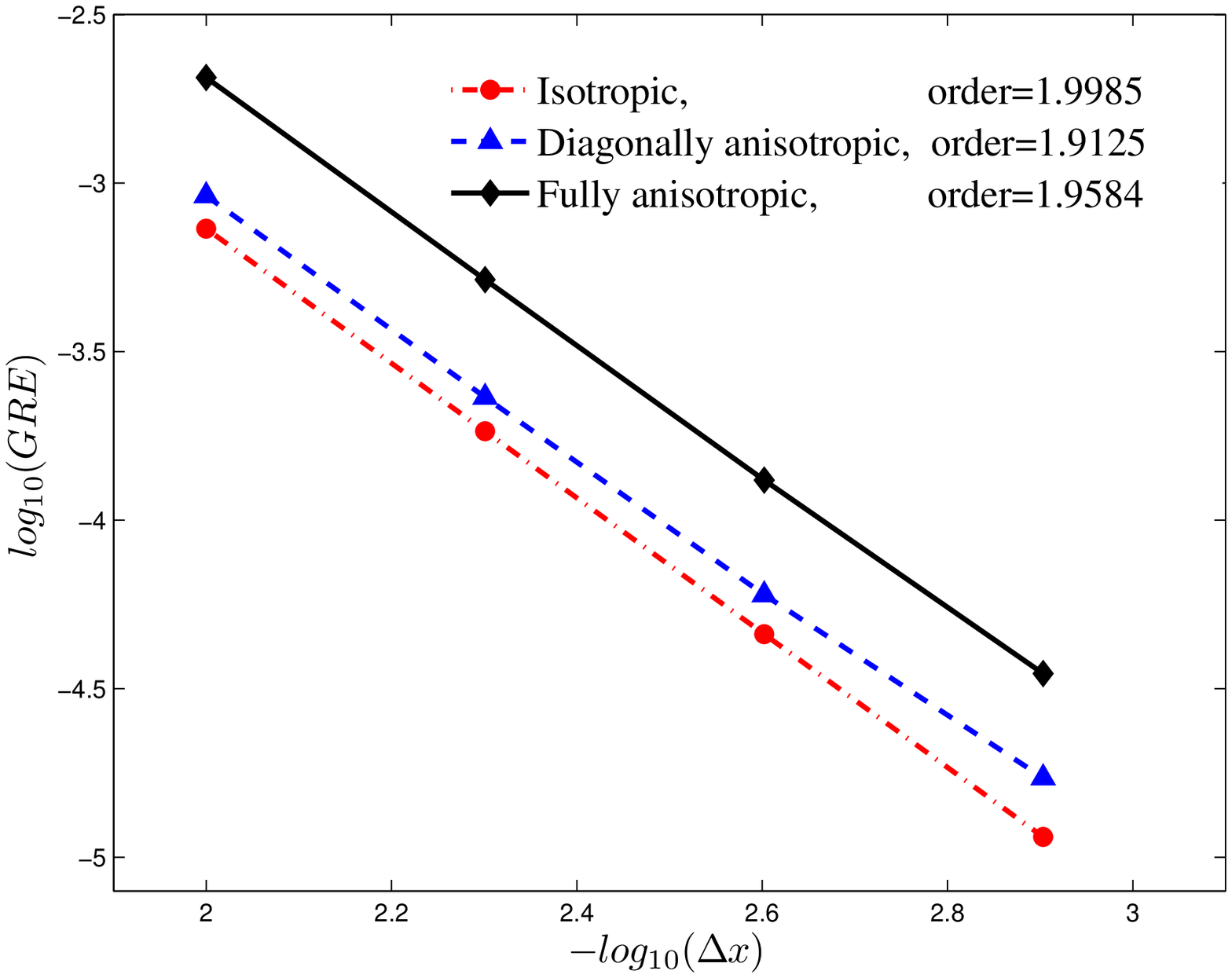}
\caption{ The global relative errors of different problems at different mesh sizes.} \label{Fig12}
\end{figure}
A number of simulations are conducted to test the capacity of present model B-TriRT model for the anisotropic CDEs, and the results at time $t=10$ are presented in Figs.~\ref{Fig9}-\ref{Fig11} where $\Delta x=1/400,~c=1.0$. As seen from these figures, one can find that the numerical solutions qualitatively agree well with analytical solutions. To give a quantitative comparison, we also calculated the $GREs$ of three types of anisotropic diffusion problem at time $t=10$, and the values of $GREs$ are $4.5926 \times 10^{-5},~6.0109 \times 10^{-5}$ and $1.3145 \times 10^{-4},$ respectively. These results indicate that the present B-TriRT model can solve anisotropic CDEs accurately. Furthermore, we tested the convergence rate of present B-TriRT model by carrying out some simulations with $\Delta  x^2/\Delta t = 5.0\times10^{-3}$ under different mesh resolutions ($\Delta x=1/100,~1/200, ~1/400,~1/800$), and the relationship between $GRE$ and lattice spacing $\Delta x$ is illustrated in Fig.~\ref{Fig12}. From this figure, one can obtain that the present B-TriRT model has a second-order accuracy in space for all types of anisotropic diffusion problems.

\subsubsection{Anisotropic convection-diffusion equation with variable diffusion tensor}
Actually, the LBGK model can also be used to solve the above Gaussian hill problem by rewriting the macroscopic equation (\ref{eq70}) into an isotropic form. For a more complicated problem with variable diffusion tensor, however, it is difficult or impossible to write it into an isotropic form. Here we consider the following problem depicted by~\cite{chai2016multiple},

\begin{equation}\label{eq73}
  \partial_t \phi+ \nabla \cdot(\phi\textbf{u})=\nabla \cdot (\textbf{A}\cdot \nabla \phi)+ S,
\end{equation}
where $\textbf{u}=(u_x,u_y)^T$ is a constant velocity, $\textbf{A}(\textbf{x},\phi)$ is a function of space $\textbf{x}=(x,y)$ and scalar variable $\phi$, and $S$ is the source term.

In our simulations, the variable diffusion tensor $\textbf{A}$ is given by

\begin{equation}\label{eq74}
  \textbf{A} =  \left(
               \begin{array}{cc}
               2-\textrm{sin}(2\pi x)\textrm{sin}(2\pi y)&0\\
                0&1
               \end{array}
             \right) \alpha,
\end{equation}
where $\alpha$ is a constant. The analytical solution of this problem is given as
\begin{equation}\label{eq76}
 \phi(x,y,t) = \textrm{exp}[(1-12\pi^2 \alpha)t]\textrm{sin}(2\pi x)\textrm{sin}(2\pi y),
\end{equation}
the source term $S$ can be expressed as
\begin{equation}\label{eq75}
\begin{aligned}
  S =& ~\textrm{exp}[(1-12\pi^{2}\alpha)t]\{\textrm{sin}(2\pi x)\textrm{sin}(2\pi y)+4\alpha\pi^2\textrm{cos}(4\pi x)\textrm{sin}^2(2\pi y)\\
   &+2\pi[u_x \textrm{cos}(2\pi x)\textrm{sin}(2\pi y)+u_y \textrm{sin}(2\pi x)\textrm{cos}(2\pi y)]\}.
  \end{aligned}
\end{equation}

\begin{figure}[ht]
\centering
\subfigure[]{ \label{fig13:a}
\includegraphics[scale=0.4]{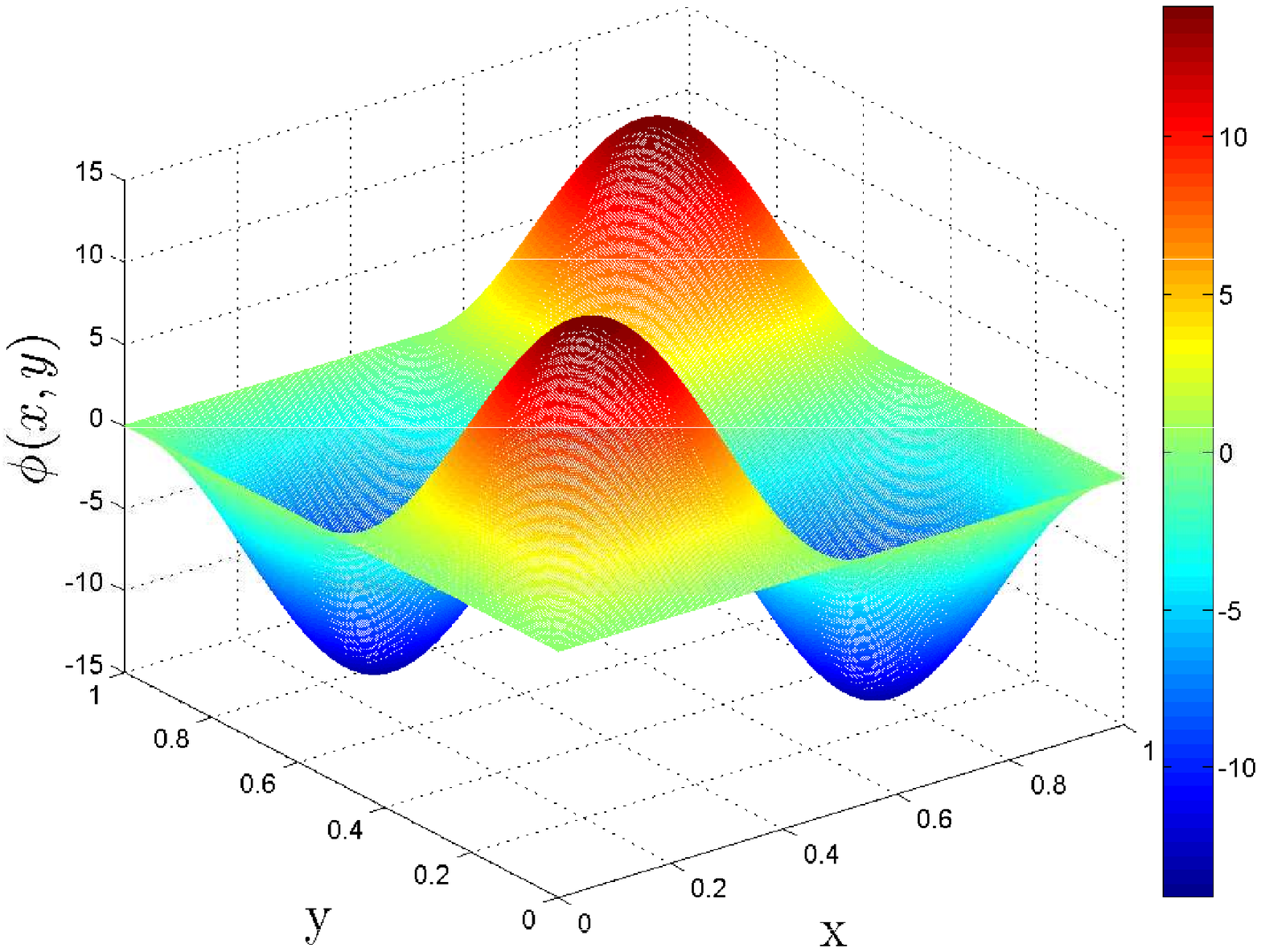}}
\subfigure[]{ \label{fig13:b}
\includegraphics[scale=0.4]{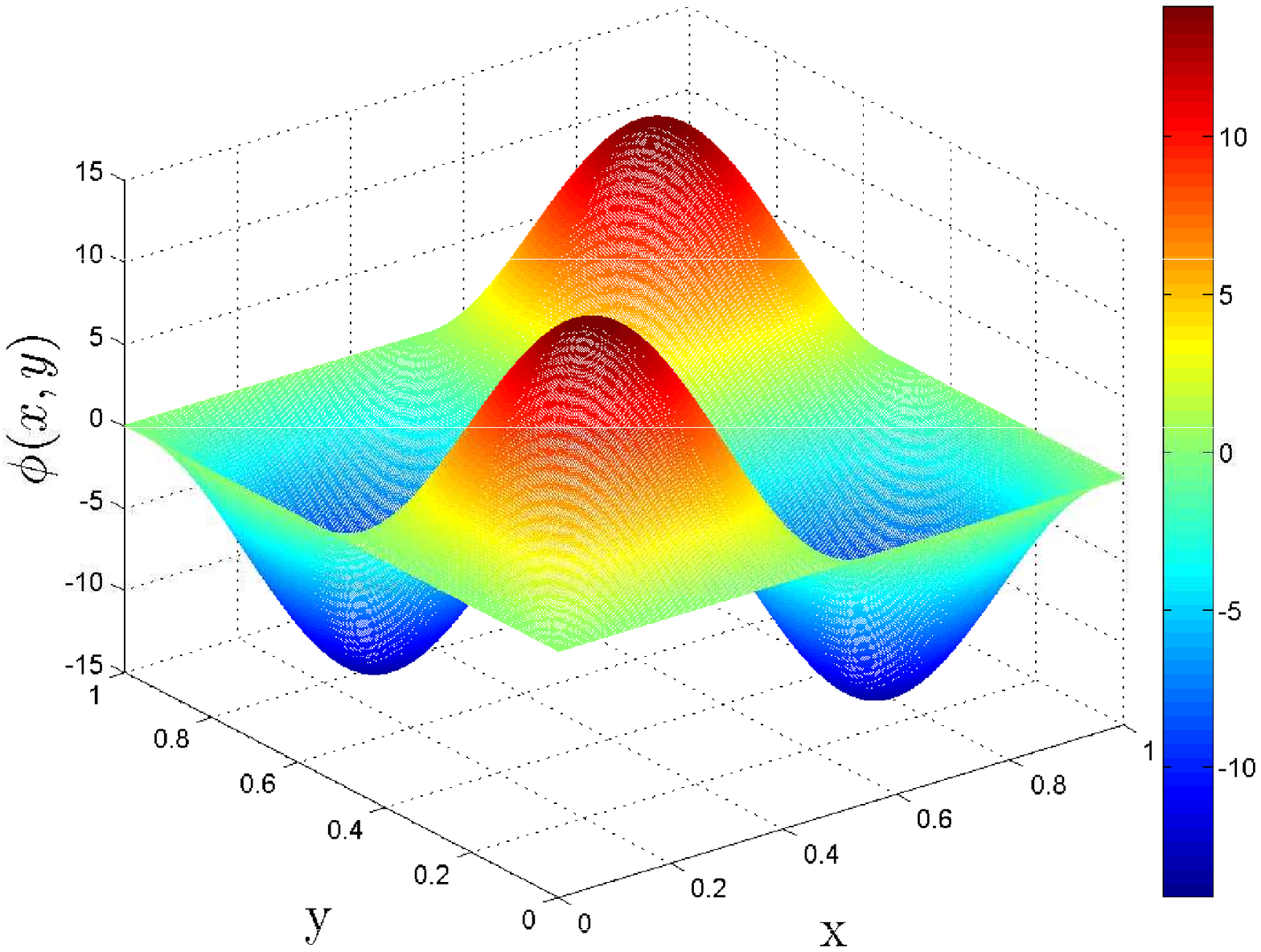}}
\caption{Numerical and analytical solutions at $t=3$[(a): numerical solution   (b): analytical solution].} \label{Fig13}
\end{figure}

\begin{figure}[ht]
\centering
\subfigure[]{ \label{fig14:a}
\includegraphics[scale=0.4]{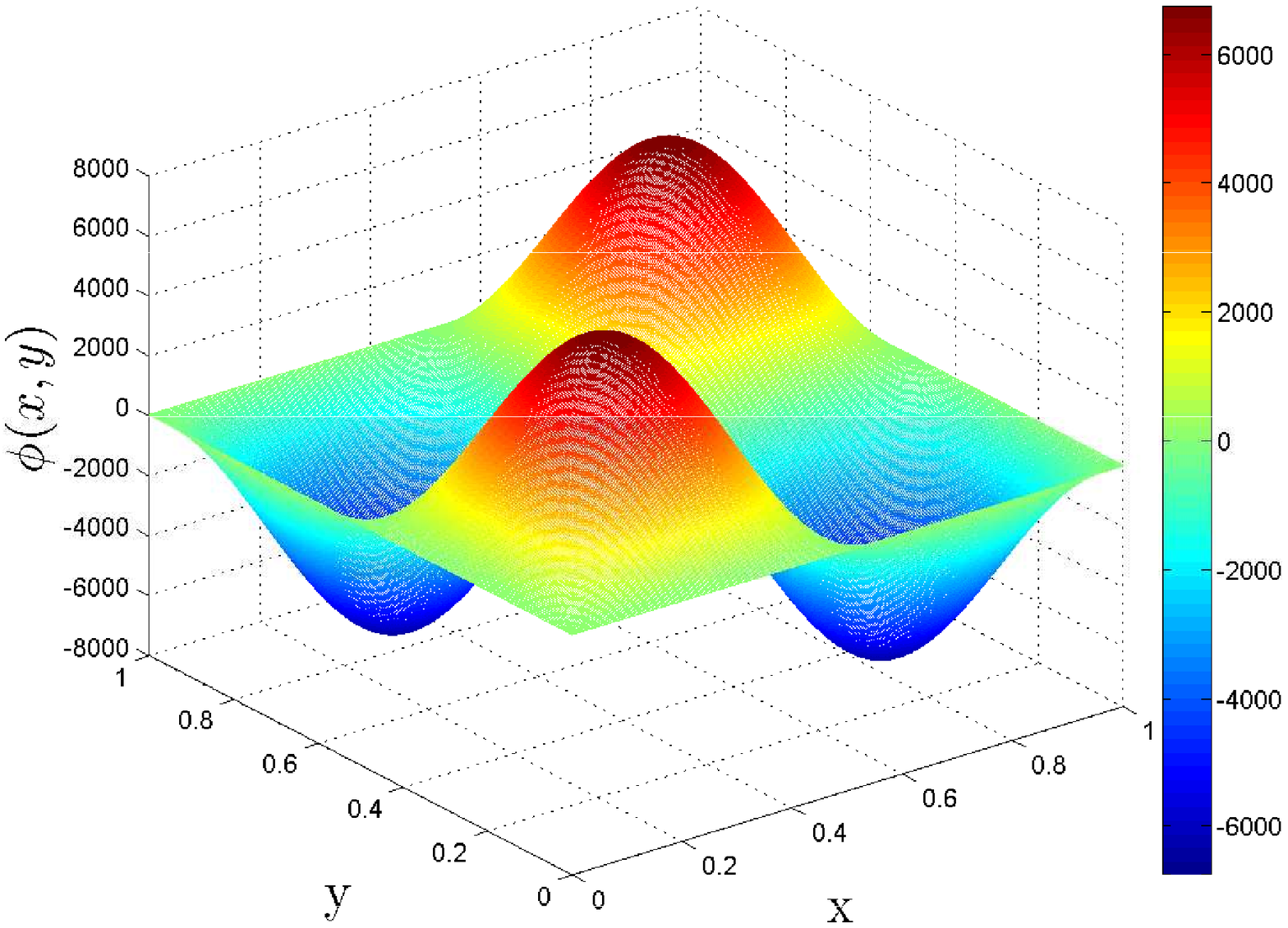}}
\subfigure[]{ \label{fig14:b}
\includegraphics[scale=0.4]{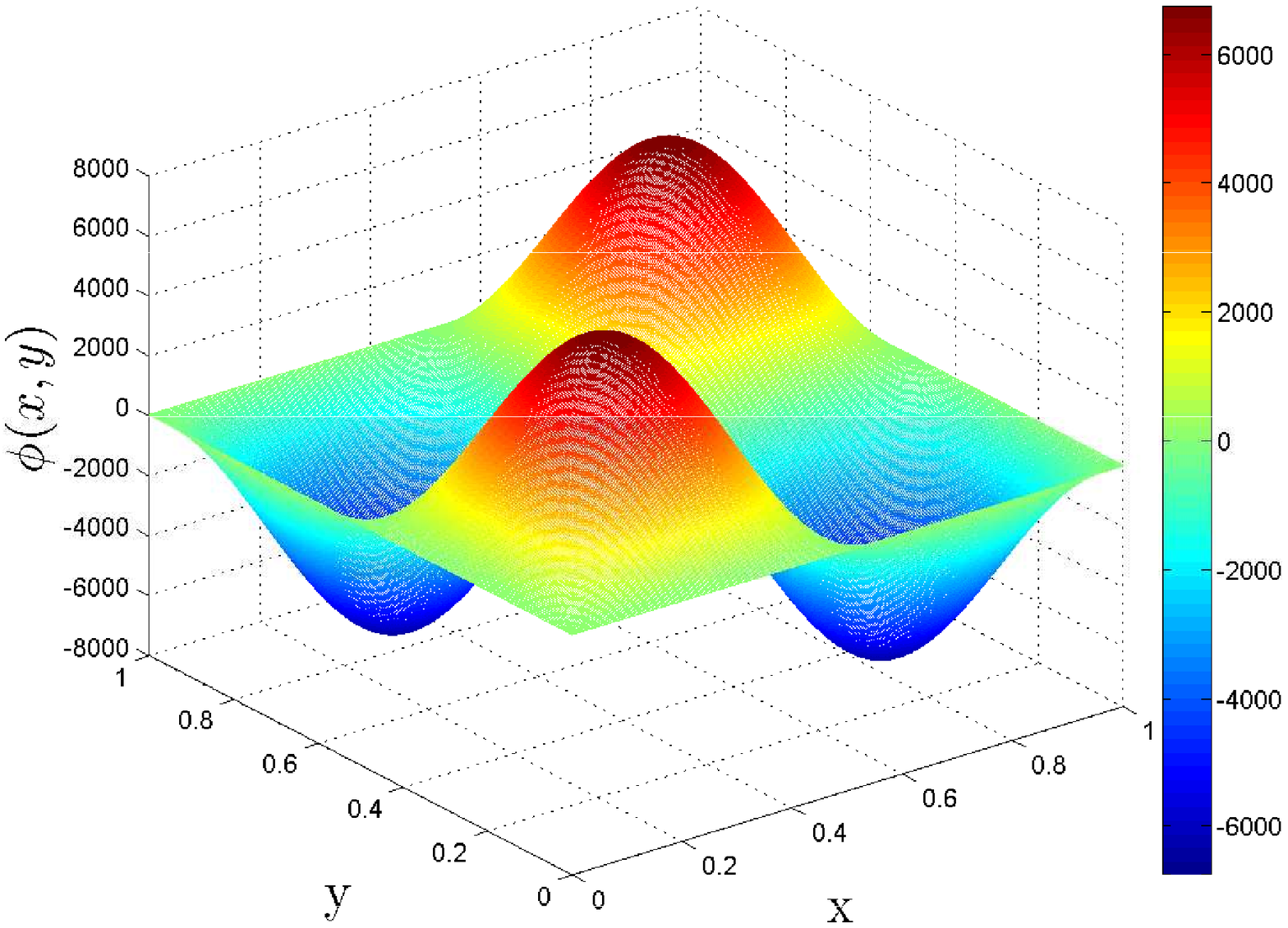}}
\caption{Numerical and analytical solutions at $t=10$ [(a): numerical solution   (b): analytical solution].} \label{Fig14}
\end{figure}

\begin{figure}[ht]
\centering
\includegraphics[scale=0.5]{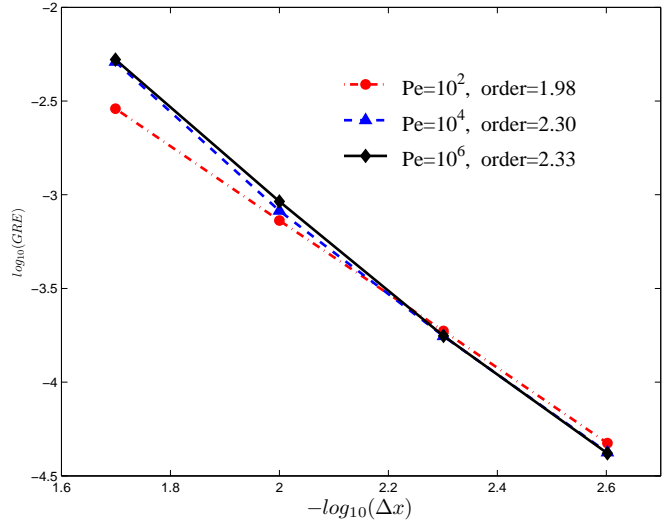}
\caption{ The global relative errors of different Pe at different mesh sizes.} \label{Fig15}
\end{figure}

We conducted some simulations with $u_x=u_y=0.1, \alpha=1.0~\times~10^{-2},~\Delta x=1/400,~c=1.0$ under different time ($t=3,~t=10$), and presented the numerical and analytical results in Figs.~\ref{Fig13}-\ref{Fig14}.  And the corresponding $GREs$ are $4.7266\times 10^{-5}$ and $4.4733\times10^{-5}.$ All the results indicate that the present B-TriRT model is accurate in solving the anisotropic CDEs with a variable diffusion tensor. In addition, we also investigated the effects of $\alpha$ under different mesh resolutions ($\Delta x=1/50,~1/100,~1/200,~1/400$), and presented the results in Tab.~\ref{tab6}. From the table, one can obtain that the present B-TriRT model is also accurate even for the case with a much smaller $\alpha.$ Based on Tab.~\ref{tab6}, we also computed the convergence rate of B-TriRT model, and the results are shown in Fig.~\ref{Fig15}. As seen from the figure, the present B-TriRT model also has a second-order convergence rate in space for this complicated anisotropic CDEs.

\begin{table}[ht]
\centering
\caption{The global relative errors of different $\alpha$ under different mesh resolutions }\label{tab6}
\begin{tabular}{c|c|c|c|c}
  \hline \hline
  $\alpha$& $\Delta x=1/50$ & $\Delta x=1/100$ & $\Delta x=1/200$ & $\Delta x=1/400$ \\
    \hline
  $1.0~\times~10^{-2}$ & $2.8801~\times~10^{-3}$  & 7.2875 $\times~ 10^{-4}$ & 1.8768 $\times~ 10^{-4}$& $4.7266~\times~10^{-5}$\\
  $1.0~\times~10^{-4}$ & $5.1174~\times~10^{-3}$ & 8.2137 $\times ~10^{-4}$ & 1.7601 $\times ~10^{-4}$ & $4.2156~\times~10^{-5}$\\
  $1.0~\times~10^{-6}$ & $5.2599~\times~10^{-3}$  & 9.2122 $\times ~10^{-4} $& 1.7660 $\times~ 10^{-4} $& $4.1732~\times~10^{-5}$\\

  \hline\hline

\end{tabular}
\end{table}

\section{Conclusions}

In this work, we presented a block triple-relaxation-time lattice Boltzmann model for general nonlinear anisotropic convection-diffusion equations, where RLBM and MLBM are its special cases. Furthermore, we  expand the non-equilibrium distribution function to the second-order moment by Hermite polynomial to obtain the present B-TriRT model.  In addition, the present B-TriRT model also has some striking features, which are summarized as follows:
 \begin{enumerate}
   \item The Chapman-Enskog analysis of present B-TriRT model is simple since it can be conducted under the LBGK framework.
   \item Based on the matrix analysis, we partitioned the relaxation parameter matrix $\textbf{S}_f$ into three relaxation parameter blocks, i.e., {\color{blue}$~\textbf{S}_0,~\textbf{S}_1,~\textbf{S}_2,$} which {\color{blue}correspond to the remaining, first-order and second-order moments of }non-equilibrium distribution function, respectively.
   \item Based on the analysis of HBB scheme for Dirichlet boundary conditions, we obtained an expression to determine the relation between relaxation parameters.
   \item The anisotropic diffusion tensor can be recovered ingeniously by relaxation parameter matrix $\textbf{K}_1$.
 \end{enumerate}
Finally, several numerical simulations (including isotropic and anisotropic CDEs) are performed, and the results show that the present model has a second-order accuracy in space, and is usually more accurate and stable than some available lattice Boltzmann models. In addition, the present B-TriRT LBM can be extended to solve the NSEs where the viscosity is related to  the relaxation matrix $\textbf{K}_2$, which would be considered in a future work.

\section{Acknowledgments}

This work is supported by the National Natural Science Foundation of China (Grants No. 51576079 and No. 51836003), and the National Key Research and Development Program of China (Grant No. 2017YFE0100100).

\section{Appendixs}
\subsection{A matrix analysis}

In this part, we will partition the relaxation parameter matrix into three relaxation parameter blocks. Now, we first rewrite the evolution Eq. (\ref{eq6}) in a vector form:
\begin{equation}\label{eqq11}
\textbf{f}^{+}=\textbf{f}-k_0\textbf{f}^{neq}- \textbf{R}\textbf{f}^{neq}- \textbf{P}\textbf{f}^{neq}+ \Dt \textbf{G}+\Dt S + \frac{\Dt^2}{2}\bar{D}S,
\end{equation}
where $\textbf{f}=\left[f_0(\textbf{x},t),f_1(\textbf{x},t),f_2(\textbf{x},t),\cdots,f_{q-1}(\textbf{x},t)\right]^T$.  The matrices $\textbf{R}$ and $\textbf{P}$ are given by
\begin{equation}\label{eqq77}
\textbf{R}_{ij}=\frac{\omega_i\textbf{c}_i\cdot[(\textbf{K}_1-k_0\textbf{I})\textbf{c}_j]}{c_s^2},~~~~\textbf{P}_{ij}=\frac{\omega_i(\textbf{c}_i\textbf{c}_i-c_s^2\textbf{I}):[(\textbf{K}_2-k_0\hat{\textbf{I}})\circ(\textbf{c}_j\textbf{c}_j)]}{2c_s^4},
\end{equation}
For simplicity, $\textbf{K}_1=k_1\textbf{I}, \textbf{K}_2=k_2\hat{\textbf{I}}$ is first considered in the following matrix analysis. Then, considering the popular D2Q9 lattice model, the explicit forms of $\textbf{R}$ and $\textbf{P}$ are given by
\begin{equation}\label{eq78}
\textbf{R}= \frac{k_1-k_0}{12}\left(
                \begin{array}{ccccccccc}
                  0 & 0 & 0 & 0 & 0 & 0 & 0 & 0 & 0 \\
                  0 & 4 & 0 & -4 & 0 & 4 & -4 & -4 & 4 \\
                  0 & 0 & 4 & 0 & -4 & 4 & 4 & -4 & -4 \\
                  0 & -4 & 0 & 4 & 0 & -4 & 4 & 4 & -4 \\
                  0 & 0 & -4 & 0 & 4 & -4 & -4 & 4 & 4 \\
                  0 & 1 & 1 & -1 & -1 & 2 & 0 & -2 & 0 \\
                  0 & -1 & 1 & 1 & -1 & 0 & 2 & 0 & -2 \\
                  0 & -1 & -1 & 1 & 1 & -2 & 0 & 2 & 0 \\
                  0 & 1 & -1 & -1 & 1 & 0 & -2 & 0 & 2 \\
                \end{array}
              \right),
\end{equation}

\begin{equation}\label{eq79}
\textbf{P}= \frac{k_2-k_0}{12}\left(
                \begin{array}{ccccccccc}
                  0 & -8 & -8 & -8 & -8 & -16 & -16 & -16 & -16 \\
                  0 & 4 & -2 & 4 & -2 & 2 & 2 & 2 & 2 \\
                  0 & -2 & 4 & -2 & 4 & 2 & 2 & 2 & 2 \\
                  0 & 4 & -2 & 4 & -2 & 2 & 2 & 2 & 2 \\
                  0 & -2 & 4 & -2 & 4 & 2 & 2 & 2 & 2 \\
                  0 & 1 & 1 & 1 & 1 & 5 & -1 & 5 & -1 \\
                  0 & 1 & 1 & 1 & 1 & -1 & 5 & -1 & 5 \\
                  0 & 1 & 1 & 1 & 1 & 5 & -1 & 5 & -1 \\
                  0 & 1 & 1 & 1 & 1 & -1 & 5 & -1 & 5 \\
                \end{array}
              \right).
\end{equation}

However, for the  D3Q19 lattice model, the matrices $\textbf{R}, \textbf{P}$ can be written as

\begin{equation}\label{eq80}
\textbf{R}= \frac{k_1-k_0}{12}\left(
                \begin{array}{ccccccccccccccccccc}
                  0  &  0  &  0  &  0  &  0 &  0 &  0 &  0 &  0 &  0 &  0 &  0 &   0 &  0 &  0 &  0 &  0 &  0 &  0\\
                  0  &  2  & -2  &  0  &  0 &  0 &  0 &  2 & -2 &  2 & -2 &  2 &  -2 &  2 & -2 &  0 &  0 &  0 &  0\\
                  0  & -2  &  2  &  0  &  0 &  0 &  0 &  2 &  2 & -2 &  2 & -2 &   2 & -2 &  2 &  0 &  0 &  0 &  0\\
                  0  &  0  &  0  &  2  & -2 &  0 &  0 &  2 &  2 & -2 & -2 &  0 &   0 &  0 &  0 &  2 & -2 &  2 & -2\\
                  0  &  0  &  0  & -2  &  2 &  0 &  0 & -2 & -2 &  2 &  2 &  0 &   0 &  0 &  0 & -2 &  2 & -2 &  2\\
                  0  &  0  &  0  &  0  &  0 &  2 & -2 &  0 &  0 &  0 &  0 &  2 &   2 & -2 & -2 &  2 &  2 & -2 & -2\\
                  0  &  0  &  0  &  0  &  0 & -2 &  2 &  0 &  0 &  0 &  0 & -2 &  -2 &  2 &  2 & -2 & -2 &  2 &  2\\
                  0  &  1  & -1  &  1  & -1 &  0 &  0 &  2 &  0 &  0 & -2 &  1 &  -1 &  1 & -1 &  1 & -1 &  1 & -1\\
                  0  & -1  &  1  &  1  & -1 &  0 &  0 &  0 &  2 & -2 &  0 & -1 &   1 & -1 &  1 &  1 & -1 &  1 & -1\\
                  0  &  1  & -1  & -1  &  1 &  0 &  0 &  0 & -2 &  2 &  0 &  1 &  -1 &  1 & -1 & -1 &  1 & -1 &  1\\
                  0  & -1  &  1  & -1  &  1 &  0 &  0 & -2 &  0 &  0 &  2 & -1 &   1 & -1 &  1 & -1 &  1 & -1 &  1\\
                  0  &  1  & -1  &  0  &  0 &  1 & -1 &  1 & -1 &  1 & -1 &  2 &   0 &  0 & -2 &  1 &  1 & -1 & -1\\
                  0  & -1  &  1  &  0  &  0 &  1 & -1 & -1 &  1 & -1 &  1 &  0 &   2 & -2 &  0 &  1 &  1 & -1 & -1\\
                  0  &  1  & -1  &  0  &  0 & -1 &  1 &  1 & -1 &  1 & -1 &  0 &  -2 &  2 &  0 & -1 & -1 &  1 &  1\\
                  0  & -1  &  1  &  0  &  0 & -1 &  1 & -1 &  1 & -1 &  1 & -2 &   0 &  0 &  2 & -1 & -1 &  1 &  1\\
                  0  &  0  &  0  &  1  & -1 &  1 & -1 &  1 &  1 & -1 & -1 &  1 &   1 & -1 & -1 &  2 &  0 &  0 & -2\\
                  0  &  0  &  0  & -1  &  1 &  1 & -1 & -1 & -1 &  1 &  1 &  1 &   1 & -1 & -1 &  0 &  2 & -2 &  0\\
                  0  &  0  &  0  &  1  & -1 & -1 &  1 &  1 &  1 & -1 & -1 & -1 &  -1 &  1 &  1 &  0 & -2 &  2 &  0\\
                  0  &  0  &  0  & -1  &  1 & -1 &  1 & -1 & -1 &  1 &  1 & -1 &  -1 &  1 &  1 & -2 &  0 &  0 &  2\\
                \end{array}
              \right),
\end{equation}

\begin{small}
\begin{equation}\label{eq81}
\textbf{P}= \frac{k_2-k_0}{24}\left(
                \begin{array}{ccccccccccccccccccc}
                  0  &-12  &-12  &-12  &-12 &-12 &-12 &-24 &-24 &-24 &-24 &-24 & -24 &-24 &-24 &-24 &-24 &-24 &-24\\
                  0  &  4  &  4  & -2  & -2 & -2 & -2 &  2 &  2 &  2 &  2 &  2 &   2 &  2 &  2 & -4 & -4 & -4 & -4\\
                  0  &  4  &  4  & -2  & -2 & -2 & -2 &  2 &  2 &  2 &  2 &  2 &   2 &  2 &  2 & -4 & -4 & -4 & -4\\
                  0  & -2  & -2  &  4  &  4 & -2 & -2 &  2 &  2 &  2 &  2 & -4 &  -4 & -4 & -4 &  2 &  2 &  2 &  2\\
                  0  & -2  & -2  &  4  &  4 & -2 & -2 &  2 &  2 &  2 &  2 & -4 &  -4 & -4 & -4 &  2 &  2 &  2 &  2\\
                  0  & -2  & -2  & -2  & -2 &  4 &  4 & -4 & -4 & -4 & -4 &  2 &   2 &  2 &  2 &  2 &  2 &  2 &  2\\
                  0  & -2  & -2  & -2  & -2 &  4 &  4 & -4 & -4 & -4 & -4 &  2 &   2 &  2 &  2 &  2 &  2 &  2 &  2\\
                  0  &  2  &  2  &  2  &  2 & -1 & -1 & 10 & -2 & -2 & 10 &  1 &   1 &  1 &  1 &  1 &  1 &  1 &  1\\
                  0  &  2  &  2  &  2  &  2 & -1 & -1 & -2 & 10 & 10 & -2 &  1 &   1 &  1 &  1 &  1 &  1 &  1 &  1\\
                  0  &  2  &  2  &  2  &  2 & -1 & -1 & -2 & 10 & 10 & -2 &  1 &   1 &  1 &  1 &  1 &  1 &  1 &  1\\
                  0  &  2  &  2  &  2  &  2 & -1 & -1 & 10 & -2 & -2 & 10 &  1 &   1 &  1 &  1 &  1 &  1 &  1 &  1\\
                  0  &  2  &  2  & -1  & -1 &  2 &  2 &  1 &  1 &  1 &  1 & 10 &  -2 & -2 & 10 &  1 &  1 &  1 &  1\\
                  0  &  2  &  2  & -1  & -1 &  2 &  2 &  1 &  1 &  1 &  1 & -2 &  10 & 10 & -2 &  1 &  1 &  1 &  1\\
                  0  &  2  &  2  & -1  & -1 &  2 &  2 &  1 &  1 &  1 &  1 & -2 &  10 & 10 & -2 &  1 &  1 &  1 &  1\\
                  0  &  2  &  2  & -1  & -1 &  2 &  2 &  1 &  1 &  1 &  1 & 10 &  -2 & -2 & 10 &  1 &  1 &  1 &  1\\
                  0  & -1  & -1  &  2  &  2 &  2 &  2 &  1 &  1 &  1 &  1 &  1 &   1 &  1 &  1& 10 &  -2 & -2 & 10\\
                  0  & -1  & -1  &  2  &  2 &  2 &  2 &  1 &  1 &  1 &  1 &  1 &   1 &  1 &  1& -2 &  10 & 10 & -2\\
                  0  & -1  & -1  &  2  &  2 &  2 &  2 &  1 &  1 &  1 &  1 &  1 &   1 &  1 &  1& -2 &  10 & 10 & -2\\
                  0  & -1  & -1  &  2  &  2 &  2 &  2 &  1 &  1 &  1 &  1 &  1 &   1 &  1 &  1& 10 &  -2 & -2 & 10\\
                \end{array}
              \right).
\end{equation}
\end{small}

As mentioned in \cite{wang2018lattice}, the matrices $\textbf{R}$ and $\textbf{P}$ can be diagonalized to a diagonal matrix with two and three entries of 1 respectively. In the previous works \cite{lallemand2000theory,chai2016multiple}, a transformation matrix \textbf{M} is employed to diagonalize the collision matrices. However, we find that this matrix cannot satisfy our need for diagonalization. Following the idea in \cite{lallemand2000theory,chai2016multiple,wang2018lattice}, one can determine a similar matrix $\textbf{M}$. To this end, we first constructed a new invertible  matrix \textbf{C} that projects populations onto natural moments $M_{opq}$ \cite{Karlin2014Gibbs}, where
\begin{equation}\label{eq82}
   M_{opq}=\langle f_i\textbf{c}_{ix}^o\textbf{c}_{iy}^p \textbf{c}_{iz}^q \rangle, o,p,q\in\{0,1,2 \},
\end{equation}
where the notation $\langle \cdots \rangle$ is used as a shorthand for summation over all the velocity indices, $d$ is the spatial dimension. From Eq. (\ref{eq82}), one can easily obtain the transformation matrix as follows,\\
in the D2Q9 lattice model,
\begin{equation}\label{eq83}
\begin{aligned}
\textbf{C}\cdot \textbf{f}=(M_{000},M_{100},M_{010},M_{200},M_{110},M_{020},M_{210},M_{120},M_{220})^{T},
\end{aligned}
\end{equation}
in the D3Q19 lattice model,
\begin{equation}\label{eq84}
\textbf{C}\cdot 
\textbf{f}=(M_{000},M_{100},M_{010},M_{001},M_{200},M_{101},M_{110},M_{011},M_{020},M_{002},M_{210},M_{201},M_{120},M_{102},M_{012},M_{021},M_{220},M_{202},M_{022})^{T},
\end{equation}

The explicit form of $\textbf{C}$ in D2Q9 and D3Q19 lattice models can be given as
\begin{equation}\label{eq85}
  \textbf{C}_{D2Q9}=\left(
  \begin{array}{ccccccccc}
    1 & 1 & 1 & 1 & 1 & 1 & 1 & 1 & 1 \\
    0 & 1 & 0 & -1& 0 & 1 & -1& -1& 1 \\
    0 & 0 & 1 & 0 & -1& 1 & 1 & -1& -1 \\
    0 & 1 & 0 & 1 & 0 & 1 & 1 & 1 & 1 \\
    0 & 0 & 0 & 0 & 0 & 1 & -1& 1 & -1 \\
    0 & 0 & 1 & 0 & 1 & 1 & 1 & 1 & 1 \\
    0 & 0 & 0 & 0 & 0 & 1 & 1 & -1& -1 \\
    0 & 0 & 0 & 0 & 0 & 1 & -1& -1& 1\\
    0 & 0 & 0 & 0 & 0 & 1 & 1 & 1 & 1 \\
  \end{array}
\right),
\end{equation}
\begin{equation}\label{eq86}
  \textbf{C}_{D3Q19}=\left(
  \begin{array}{ccccccccccccccccccc}
1	&1	&1	&1	&1	&1	&1	&1	&1	&1	&1	&1	&1	&1	&1	&1	&1	&1	&1\\
0	&1	&-1	&0	&0	&0	&0	&1	&-1	&1	&-1	&1	&-1	&1	&-1	&0	&0	&0	&0\\
0	&0	&0	&1	&-1	&0	&0	&1	&1	&-1	&-1	&0	&0	&0	&0	&1	&-1	&1	&-1\\
0	&0	&0	&0	&0	&1	&-1	&0	&0	&0	&0	&1	&1	&-1	&-1	&1	&1	&-1	&-1\\
0	&1	&1	&0	&0	&0	&0	&1	&1	&1	&1	&1	&1	&1	&1	&0  &0	&0	&0\\
0	&0	&0	&0	&0	&0	&0	&0	&0	&0	&0	&1	&-1	&-1	&1	&0	&0	&0	&0\\
0	&0	&0	&0	&0	&0	&0	&1	&-1	&-1	&1	&0	&0	&0	&0	&0	&0	&0	&0\\
0	&0	&0	&0	&0	&0	&0	&0	&0	&0	&0	&0	&0	&0	&0	&1	&-1	&-1	&1\\
0	&0	&0	&1	&1	&0	&0	&1	&1	&1	&1	&0	&0	&0	&0	&1	&1	&1	&1\\
0	&0	&0	&0	&0	&1	&1	&0	&0	&0	&0	&1	&1	&1	&1	&1	&1	&1	&1\\
0	&0	&0	&0	&0	&0	&0	&1	&1	&-1	&-1	&0	&0	&0	&0	&0	&0	&0	&0\\
0	&0	&0	&0	&0	&0	&0	&0	&0	&0	&0	&1	&1	&-1	&-1	&0	&0	&0	&0\\
0	&0	&0	&0	&0	&0	&0	&1	&-1	&1	&-1	&0	&0	&0	&0	&0	&0	&0	&0\\
0	&0	&0	&0	&0	&0	&0	&0	&0	&0	&0	&1	&-1	&1	&-1	&0	&0  &0	&0\\
0	&0	&0	&0	&0	&0	&0	&0	&0	&0	&0	&0	&0	&0	&0	&1	&-1	&1	&-1\\
0	&0	&0	&0	&0	&0	&0	&0	&0	&0	&0	&0	&0	&0	&0	&1	&1	&-1	&-1\\
0	&0	&0	&0	&0	&0	&0	&1	&1	&1	&1	&0	&0	&0	&0	&0	&0	&0	&0\\
0	&0	&0	&0	&0	&0	&0	&0	&0	&0	&0	&1	&1	&1	&1	&0	&0	&0	&0\\
0	&0	&0	&0	&0	&0	&0	&0	&0	&0	&0	&0	&0	&0	&0	&1	&1	&1	&1\\
  \end{array}
\right).
\end{equation}
We note that matrix $\textbf{C}$ cannot diagonalize the matrix $\textbf{R}$ or $\textbf{P}$. However, through an elementary transformation
\begin{equation}\label{eq87}
    \textbf{T}=\textbf{H}\textbf{C}
\end{equation}
where
\begin{equation}\label{eq88}
  \textbf{T}_{D2Q19}=\left(
  \begin{array}{ccccccccc}
    1 & 1 & 1 & 1 & 1 & 1 & 1 & 1 & 1 \\
    0 & 1 & 0 & -1 & 0 & 1 & -1 & -1 & 1 \\
    0 & 0 & 1 & 0 & -1 & 1 & 1 & -1 & -1 \\
    0 & 1 & 0 & 1 & 0 & 1 & 1 & 1 & 1 \\
    0 & 0 & 0 & 0 & 0 & 1 & -1 & 1 & -1 \\
    0 & 0 & 1 & 0 & 1 & 1 & 1 & 1 & 1 \\
    0 & 0 & -\frac{1}{3} & 0 & \frac{1}{3} & \frac{2}{3} & \frac{2}{3} & -\frac{2}{3} & -\frac{2}{3} \\
    0 & -\frac{1}{3} & 0 & \frac{1}{3} & 0 & \frac{2}{3} & -\frac{2}{3} & -\frac{2}{3} & \frac{2}{3} \\
    0 & -\frac{1}{3} & -\frac{1}{3} & -\frac{1}{3} & -\frac{1}{3} & \frac{1}{3} & \frac{1}{3} & \frac{1}{3} & \frac{1}{3} \\
  \end{array}
\right),
\end{equation}

\begin{equation}\label{eq89}
  \textbf{T}_{D3Q19}=\left(
  \begin{array}{ccccccccccccccccccc}
1    	&1    	&1    	&1    	&1    	&1    	&1    	&1    	&1    	&1    	&1    	&1    	&1    	&1    	&1    	&1    	&1    	&1    	&1\\
0    	&1    	&-1    	&0    	&0    	&0    	&0    	&1    	&-1    	&1    	&-1    	&1    	&-1    	&1    	&-1    	&0    	&0    	&0    	&0\\
0    	&0    	&0    	&1    	&-1    	&0    	&0    	&1    	&1    	&-1    	&-1    	&0    	&0    	&0    	&0    	&1    	&-1    	&1    	&-1\\
0    	&0    	&0    	&0    	&0    	&1    	&-1    	&0    	&0    	&0    	&0    	&1    	&1    	&-1    	&-1    	&1    	&1    	&-1    	&-1\\
0    	&1    	&1    	&0    	&0    	&0    	&0    	&1    	&1    	&1    	&1    	&1    	&1    	&1    	&1    	&0    	&0    	&0    	&0\\
0    	&0    	&0    	&0    	&0    	&0    	&0    	&0    	&0    	&0    	&0    	&1    	&-1    	&-1    	&1    	&0    	&0    	&0    	&0\\
0    	&0    	&0    	&0    	&0    	&0    	&0    	&1    	&-1    	&-1    	&1    	&0    	&0    	&0    	&0    	&0    	&0    	&0    	&0\\
0    	&0    	&0    	&0    	&0    	&0    	&0    	&0    	&0    	&0    	&0    	&0    	&0    	&0    	&0    	&1    	&-1    	&-1    	&1\\
0    	&0    	&0    	&1    	&1    	&0    	&0    	&1    	&1    	&1    	&1    	&0    	&0    	&0    	&0    	&1    	&1    	&1    	&1\\
0    	&0    	&0    	&0    	&0    	&1    	&1    	&0    	&0    	&0    	&0    	&1    	&1    	&1    	&1    	&1    	&1    	&1    	&1\\
0    	&0    	&0    	&- \frac{1}{3}	 &\frac{1}{3}	&0    	&0    	 &\frac{2}{3}	 &\frac{2}{3}	&- \frac{2}{3}	&- \frac{2}{3}	&0    	&0    	&0    	&0    	&- \frac{1}{3} &\frac{1}{3}&- \frac{1}{3}	 &\frac{1}{3}\\
0    	&0    	&0    	&0    	&0    	&- \frac{1}{3}	 &\frac{1}{3}	&0    	&0    	&0    	&0    	 &\frac{2}{3}	 &\frac{2}{3}	&- \frac{2}{3}	&- \frac{2}{3}	&- \frac{1}{3}	&- \frac{1}{3} &\frac{1}{3}	 &\frac{1}{3}\\
0    	&- \frac{1}{3}	 &\frac{1}{3}	&0    	&0    	&0    	&0    	 &\frac{2}{3}	&- \frac{2}{3}	 &\frac{2}{3}	&- \frac{2}{3}	&- \frac{1}{3}	 &\frac{1}{3}	&- \frac{1}{3}	 &\frac{1}{3}	&0    	&0    	&0    	&0\\
0    	&- \frac{1}{3}	 &\frac{1}{3}	&0    	&0    	&0    	&0    	&- \frac{1}{3}	 &\frac{1}{3}	&- \frac{1}{3}	 &\frac{1}{3}	 &\frac{2}{3}	&- \frac{2}{3}	 &\frac{2}{3}	&- \frac{2}{3}	&0    	&0    	&0    	&0\\
0    	&0    	&0    	&- \frac{1}{3}	 &\frac{1}{3}	&0    	&0    	&- \frac{1}{3}	&- \frac{1}{3}	 &\frac{1}{3}	 &\frac{1}{3}	&0    	&0    	&0    	&0    	 &\frac{2}{3}	&- \frac{2}{3}	 &\frac{2}{3}	&- \frac{2}{3}\\
0    	&0    	&0    	&0    	&0    	&- \frac{1}{3}	 &\frac{1}{3}	&0    	&0    	&0    	&0    	&- \frac{1}{3}	&- \frac{1}{3}	 &\frac{1}{3}	 &\frac{1}{3}	 &\frac{2}{3}	 &\frac{2}{3}	&- \frac{2}{3}	&- \frac{2}{3}\\
0    	&- \frac{1}{3}	&- \frac{1}{3}	&- \frac{1}{3}	&- \frac{1}{3}	&\frac{1}{6}   	&\frac{1}{6}  	 &\frac{1}{3}	 &\frac{1}{3}	 &\frac{1}{3}	 &\frac{1}{3}	&- \frac{1}{6}	&- \frac{1}{6}&- \frac{1}{6}	&- \frac{1}{6}	&- \frac{1}{6}	&- \frac{1}{6}	&- \frac{1}{6}	&- \frac{1}{6}\\
0    	&- \frac{1}{3}	&- \frac{1}{3}	&\frac{1}{6}  	&\frac{1}{6}    	&- \frac{1}{3}	&- \frac{1}{3}	&- \frac{1}{6} 	&- \frac{1}{6} 	&- \frac{1}{6} 	&- \frac{1}{6} 	 &\frac{1}{3}	 &\frac{1}{3}	 &\frac{1}{3}	 &\frac{1}{3}	&- \frac{1}{6} 	&- \frac{1}{6} 	&- \frac{1}{6} 	&- \frac{1}{6} \\
0    	&\frac{1}{6}    	&\frac{1}{6}   	&- \frac{1}{3}	&- \frac{1}{3}	&- \frac{1}{3}	&- \frac{1}{3}	&- \frac{1}{6}	&- \frac{1}{6}	&- \frac{1}{6}	&- \frac{1}{6}	&- \frac{1}{6}	&- \frac{1}{6}	&- \frac{1}{6}	&- \frac{1}{6}	 &\frac{1}{3}	 &\frac{1}{3} &\frac{1}{3} &\frac{1}{3}

\end{array}
\right).
\end{equation}

\begin{equation}\label{eq90}
  \textbf{H}_{D2Q9}=\left(
               \begin{array}{ccccccccc}
                 1 & 0 & 0 & 0 & 0 & 0 & 0 & 0 & 0 \\
                 0 & 1 & 0 & 0 & 0 & 0 & 0 & 0 & 0 \\
                 0 & 0 & 1 & 0 & 0 & 0 & 0 & 0 & 0 \\
                 0 & 0 & 0 & 1 & 0 & 0 & 0 & 0 & 0 \\
                 0 & 0 & 0 & 0 & 1 & 0 & 0 & 0 & 0 \\
                 0 & 0 & 0 & 0 & 0 & 1 & 0 & 0 & 0 \\
                 0 & 0 & -\frac{1}{3} & 0 & 0 & 0 & 1 & 0 & 0 \\
                 0 & -\frac{1}{3} & 0 & 0 & 0 & 0 & 0 & 1 & 0 \\
                 0 & 0 & 0 & -\frac{1}{3}& 0 & -\frac{1}{3} & 0 & 0 & 1 \\
               \end{array}
             \right),
\end{equation}

\begin{equation}\label{eq91}
  \textbf{H}_{D3Q19}=\left(
               \begin{array}{ccccccccccccccccccc}
                 1 & 0 & 0 & 0 & 0 & 0 & 0 & 0 & 0 & 0 & 0 & 0 & 0 & 0 & 0 & 0 & 0 & 0 & 0  \\
                 0 & 1 & 0 & 0 & 0 & 0 & 0 & 0 & 0 & 0 & 0 & 0 & 0 & 0 & 0 & 0 & 0 & 0 & 0  \\
                 0 & 0 & 1 & 0 & 0 & 0 & 0 & 0 & 0 & 0 & 0 & 0 & 0 & 0 & 0 & 0 & 0 & 0 & 0  \\
                 0 & 0 & 0 & 1 & 0 & 0 & 0 & 0 & 0 & 0 & 0 & 0 & 0 & 0 & 0 & 0 & 0 & 0 & 0  \\
                 0 & 0 & 0 & 0 & 1 & 0 & 0 & 0 & 0 & 0 & 0 & 0 & 0 & 0 & 0 & 0 & 0 & 0 & 0  \\
                 0 & 0 & 0 & 0 & 0 & 1 & 0 & 0 & 0 & 0 & 0 & 0 & 0 & 0 & 0 & 0 & 0 & 0 & 0  \\
                 0 & 0 & 0 & 0 & 0 & 0 & 1 & 0 & 0 & 0 & 0 & 0 & 0 & 0 & 0 & 0 & 0 & 0 & 0  \\
                 0 & 0 & 0 & 0 & 0 & 0 & 0 & 1 & 0 & 0 & 0 & 0 & 0 & 0 & 0 & 0 & 0 & 0 & 0  \\
                 0 & 0 & 0 & 0 & 0 & 0 & 0 & 0 & 1 & 0 & 0 & 0 & 0 & 0 & 0 & 0 & 0 & 0 & 0  \\
                 0 & 0 & 0 & 0 & 0 & 0 & 0 & 0 & 0 & 1 & 0 & 0 & 0 & 0 & 0 & 0 & 0 & 0 & 0  \\
                 0 & 0 & -\frac{1}{3} & 0 & 0 & 0 & 0 & 0 & 0 & 0 & 1 & 0 & 0 & 0 & 0 & 0 & 0 & 0 & 0  \\
                 0 & 0 & 0 & -\frac{1}{3} & 0 & 0 & 0 & 0 & 0 & 0 & 0 & 1 & 0 & 0 & 0 & 0 & 0 & 0 & 0  \\
                 0 & -\frac{1}{3} & 0 & 0 & 0 & 0 & 0 & 0 & 0 & 0 & 0 & 0 & 1 & 0 & 0 & 0 & 0 & 0 & 0  \\
                 0 & -\frac{1}{3} & 0 & 0 & 0 & 0 & 0 & 0 & 0 & 0 & 0 & 0 & 0 & 1 & 0 & 0 & 0 & 0 & 0  \\
                 0 & 0 & -\frac{1}{3} & 0 & 0 & 0 & 0 & 0 & 0 & 0 & 0 & 0 & 0 & 0 & 1 & 0 & 0 & 0 & 0  \\
                 0 & 0 & 0 & -\frac{1}{3} & 0 & 0 & 0 & 0 & 0 & 0 & 0 & 0 & 0 & 0 & 0 & 1 & 0 & 0 & 0  \\
                 0 & 0 & 0 & 0 & -\frac{1}{3} & 0 & 0 & 0 & -\frac{1}{3} & \frac{1}{6} & 0 & 0 & 0 & 0 & 0 & 0 & 1 & 0 & 0  \\
                 0 & 0 & 0 & 0 & -\frac{1}{3} & 0 & 0 & 0 & \frac{1}{6} & -\frac{1}{3} & 0 & 0 & 0 & 0 & 0 & 0 & 0 & 1 & 0  \\
                 0 & 0 & 0 & 0 & \frac{1}{6} & 0 & 0 & 0 & -\frac{1}{3} & -\frac{1}{3} & 0 & 0 & 0 & 0 & 0 & 0 & 0 & 0 & 1  \\
               \end{array}
             \right),
\end{equation}
we could obtain the common transformation matrix $\textbf{T}$ to diagonalize the matrices $\textbf{R}$ and $\textbf{P}$ as follows:\\
in the D2Q9 lattice model,\\
\begin{equation}\label{eq92}
\begin{aligned}
\textbf{T}\textbf{R}\textbf{T}^{-1}=\textbf{S}_{R}&=\mathrm{diag}(0,k_1-k_0,k_1-k_0,0,0,0,0,0,0),\\
\textbf{T}\textbf{P}\textbf{T}^{-1}=\textbf{S}_{P}&=\mathrm{diag}(0,0,0,k_2-k_0,k_2-k_0,k_2-k_0,0,0,0),
\end{aligned}
\end{equation}
in the D3Q19 lattice model,\\
\begin{equation}\label{eq93}
\begin{aligned}
\textbf{T}\textbf{R}\textbf{T}^{-1}=\textbf{S}_{R}&=\mathrm{diag}(0,k_1-k_0,k_1-k_0,k_1-k_0,0,0,0,0,0,0,0,0,0,0,0,0,0,0,0),\\
\textbf{T}\textbf{P}\textbf{T}^{-1}=\textbf{S}_{P}&=\mathrm{diag}(0,0,0,0,k_2-k_0,k_2-k_0,k_2-k_0,k_2-k_0,k_2-k_0,k_2-k_0,0,0,0,0,0,0,0,0,0).
\end{aligned}
\end{equation}

Based on Eq.~(\ref{eq23}), the evolution Eq.~(\ref{eq10}) can be rewritten as
\begin{equation}\label{eq94}
  f_i(\textbf{x}+\textbf{c}_i\Delta t,t+\Delta t)=f_i(\textbf{x},t)-(\textbf{T}^{-1}\textbf{S}_f\textbf{T})_{ij}f_j^{neq}+\Dt G_i+\Dt S_i + \frac{\Dt^2}{2}\bar{D}_iS_i,
\end{equation}
where $\textbf{S}_f$ is defined as\\
\begin{equation}\label{eq95}
 \mathrm{D2Q9}: \textbf{S}_f=k_0\textbf{I}_{9\times9}+\textbf{S}_{R}+\textbf{S}_{P}=\mathrm{diag}(k_0,k_1,k_1,k_2,k_2,k_2,k_0,k_0,k_0);
\end{equation}
\begin{equation}\label{eq96}
 \mathrm{D3Q19}: \textbf{S}_f=k_0\textbf{I}_{19\times19}+\textbf{S}_{R}+\textbf{S}_{P}=\mathrm{diag}(k_0,k_1,k_1,k_1,k_2,k_2,k_2,k_2,k_2,k_2,k_0,k_0,k_0,k_0,k_0,k_0,k_0,k_0,k_0).
\end{equation}
In addition, for general relaxation parameter blocks $\textbf{K}_1,~\textbf{K}_2$ being off-diagonal matrices as
\begin{equation}\label{eq98}
  \mathrm{D2Q9}: \textbf{K}_1=\left(
               \begin{array}{cc}
                k_{1,11}&k_{1,12}\\
                k_{1,21}&k_{1,22}
               \end{array}
             \right),
    \textbf{K}_2=\left(
               \begin{array}{cc}
                k_{2,11}&k_{2,12}\\
                k_{2,21}&k_{2,22}
               \end{array}
             \right),
\end{equation}

\begin{equation}\label{eq101}
  \mathrm{D3Q19}: \textbf{K}_1=\left(
               \begin{array}{ccc}
                k_{1,11}&k_{1,12}&k_{1,13}\\
                k_{1,21}&k_{1,22}&k_{1,23}\\
                {\color{blue}k_{1,31}}&k_{1,32}&k_{1,33}
               \end{array}
             \right),~~~~
    \textbf{K}_2=\left(
               \begin{array}{ccc}
                k_{2,11}&k_{2,12}&k_{2,13}\\
                k_{2,21}&k_{2,22}&k_{2,23}\\
               {\color{blue}k_{3,31}} &k_{2,32}&k_{2,33}
               \end{array}
             \right),
\end{equation}

the relaxation parameter matrix will become a block matrix, and can be given as
\begin{equation}\label{eq99}
 \mathrm{D2Q9}: \textbf{S}_f=\left(
               \begin{array}{ccccccccc}
 k_0 &   0&   0&     0&             0&     0&  0&  0&  0\\
   0& k_{1,11}& k_{1,12}&     0&             0&     0&  0&  0&  0\\
  0& k_{1,21}& k_{1,22}&     0&             0&     0&  0&  0&  0\\
  0&   0&   0&   k_{2,11}&    0       &     0&  0&  0&  0\\
  0&   0&   0& {\color{blue}0} & \frac{k_{2,12}}{2} + \frac{k_{2,21}}{2}& 0&  0&  0&  0\\
  0&   0&   0&     0&     0      &   k_{2,22}&  0&  0&  0\\
  0&   0&   0&     0&             0&     0& k_0&  0&  0\\
  0&   0&   0&     0&             0&     0&  0& k_0&  0\\
  0&   0&   0&     0&             0&     0&  0&  0& k_0
               \end{array}
 \right),
\end{equation}

D3Q19:\\
\begin{equation}\label{eq100}
  \textbf{S}_f=\left(
               \begin{array}{ccccccccccccccccccc}
 k_0&       0&        0&        0&   0&             0&             0&             0&   0&   0&  0&  0&  0&  0&  0&  0&  0&  0&  0\\
  0& k_{1,11}& k_{1,12}& k_{1,13}&   0&             0&             0&             0&   0&   0&  0&  0&  0&  0&  0&  0&  0&  0&  0\\
  0&k_{1,21} & k_{1,22}& k_{1,23}&   0&             0&             0&             0&   0&   0&  0&  0&  0&  0&  0&  0&  0&  0&  0\\
  0&k_{1,31} &k_{1,32} & k_{1,33}&   0&             0&             0&             0&   0&   0&  0&  0&  0&  0&  0&  0&  0&  0&  0\\
  0&        0&        0&        0& k_{2,11}&        0&             0&             0&   0&   0&  0&  0&  0&  0&  0&  0&  0&  0&  0\\
  0&        0&        0&        0&   0& \frac{k_{2,13}}{2} + \frac{k_{2,31}}{2}&             0&             0&   0&   0&  0&  0&  0&  0&  0&  0&  0&  0&  0\\
  0&        0&        0&        0&   0&             0& \frac{k_{2,12}}{2} + \frac{k_{2,21}}{2}&             0&   0&   0&  0&  0&  0&  0&  0&  0&  0&  0&  0\\
  0&        0&        0&        0&   0&             0&             0& \frac{k_{2,23}}{2} + \frac{k_{2,32}}{2}&   0&   0&  0&  0&  0&  0&  0&  0&  0&  0&  0\\
  0&        0&        0&        0&   0&             0&             0&             0& k_{2,22}&   0&  0&  0&  0&  0&  0&  0&  0&  0&  0\\
  0&        0&        0&        0&   0&             0&             0&             0&   0& k_{2,33}&  0&  0&  0&  0&  0&  0&  0&  0&  0\\
  0&        0&        0&        0&   0&             0&             0&             0&   0&   0& k_0&  0&  0&  0&  0&  0&  0&  0&  0\\
  0&        0&        0&        0&   0&             0&             0&             0&   0&   0&  0& k_0&  0&  0&  0&  0&  0&  0&  0\\
  0&        0&        0&        0&   0&             0&             0&             0&   0&   0&  0&  0& k_0&  0&  0&  0&  0&  0&  0\\
  0&        0&        0&        0&   0&             0&             0&             0&   0&   0&  0&  0&  0& k_0&  0&  0&  0&  0&  0\\
  0&        0&        0&        0&   0&             0&             0&             0&   0&   0&  0&  0&  0&  0& k_0&  0&  0&  0&  0\\
  0&        0&        0&        0&   0&             0&             0&             0&   0&   0&  0&  0&  0&  0&  0& k_0&  0&  0&  0\\
  0&        0&        0&        0&   0&             0&             0&             0&   0&   0&  0&  0&  0&  0&  0&  0& k_0&  0&  0\\
  0&        0&        0&        0&   0&             0&             0&             0&   0&   0&  0&  0&  0&  0&  0&  0&  0& k_0&  0\\
  0&        0&        0&        0&   0&             0&             0&             0&   0&   0&  0&  0&  0&  0&  0&  0&  0&  0& k_0\\
               \end{array}
 \right).
\end{equation}

Finally, based on Eqs.(~\ref{eq92}),~(\ref{eq93}),~(\ref{eq99}), and (\ref{eq100}), one can find that $\textbf{S}_f$  could be partitioned into several relaxation parameter blocks as
\begin{equation}\label{eqqqq}
 \textbf{S}_f=\mathrm{diag}(\textbf{S}_0, \textbf{S}_1, \textbf{S}_2, \cdots, \textbf{S}_m),~~~~ m<q,
\end{equation}
where $m$ represents the order of moment, $q$ represents the number of discrete velocies. Moreover, putting the relaxation parameters of zero-order, thirdly-order and higher-order moments together,  we could partition the relaxation parameter matrix $\textbf{S}_f$ into three blocks,
\begin{equation}\label{eq97}
  \textbf{S}_f=\mathrm{diag}(\textbf{S}_0,~\textbf{S}_1,~\textbf{S}_2),
\end{equation}
where $\textbf{S}_1,~\textbf{S}_2,~\textbf{S}_0,$ correspond to the first-order, second-order moment and remaining moments of non-equilibrium distribution function, respectively.

\section{References}
\bibliographystyle{elsarticle-num}
\bibliography{ref}
\end{document}